\documentclass[aps,prb,twocolumn, amsfonts,showpacs,longbibliography,citeautoscript,superscriptaddress,]{revtex4-2}
\usepackage[toc,page]{appendix}
\usepackage[british]{babel}
\usepackage[utf8]{inputenc}
\usepackage{amsmath,amssymb}
\usepackage{graphicx}
\usepackage{ae}
\usepackage{esdiff}
\usepackage{braket}
\usepackage{bbm}
\usepackage{simplewick}
\usepackage{float}
\usepackage{braket}
\usepackage{simplewick}
\usepackage{color}
\usepackage{framed}
\usepackage[caption=false]{subfig}
\usepackage[pdftex,colorlinks=true,linkcolor=blue,citecolor=blue,urlcolor=black]{hyperref}
\usepackage{hyperref}
\usepackage{mathtools}
\usepackage{enumitem}
\usepackage{ragged2e}
\usepackage{floatflt}
\usepackage{import}
\usepackage{titlesec}
\usepackage{multirow}
\usepackage{etoolbox}
\usepackage{appendix}
\usepackage{dsfont}
\usepackage{comment}
\usepackage{xcolor}
\usepackage{blindtext}
\usepackage[makeroom]{cancel}
\usepackage{mathtools}
\begin{document}

\newcommand{\norm}[1]{\left\lVert#1\right\rVert}

\def \r{{\boldsymbol{r}}}
\def \lvec{{\boldsymbol{l}}}
\def \x{{\boldsymbol{x}}}
\def \k{{\boldsymbol{k}}}
\def \p{{\boldsymbol{p}}}
\def \q{{\boldsymbol{q}}}
\def \dl{\frac{\partial}{\partial l}}
\def \P{{\boldsymbol{P}}}
\def \K{{\boldsymbol{K}}}
\def \piph{\Pi_\text{ph}}
\def \sign{ \text{sign}}
\def \lamt{\tilde{\lambda}}
\def \intk{{\int_\textbf{k}}}
\def \Ims{\text{Im} [ \Sigma^D(\omega) ]}
\def \Gammabs{\Gamma_{\text{bs}}}
\def \aq{|\q|}
\def \ak{|\k|}
\def \Vt{\tilde V}
\def \omp{\omega^\prime}
\def \om{\omega_m}
\def \on{\omega_n}
\def \omt{\tilde{\omega}}
\def \T{\mathcal{T}}

\definecolor{mgrey}{RGB}{63,63,63}
\definecolor{mred}{RGB}{235,97,51}
\newcommand{\mg}[1]{{\color{mgrey}{#1}}}
\newcommand{\mr}[1]{{\color{mred}{#1}}}

\newcommand{\red}[1]{{\color{red}{#1}}}
\newcommand{\blue}[1]{{\color{blue}{#1}}}
\newcommand{\al}[1]{\begin{align}#1\end{align}}
\newcommand{\beq} {\begin{equation}}
\newcommand{\eeq} {\end{equation}}
\newcommand{\bea} {\begin{eqnarray}}
\newcommand{\eea} {\end{eqnarray}}
\newcommand{\be} {\begin{equation}}
\newcommand{\ee} {\end{equation}}

\def\BigColSep{\setlength{\arraycolsep}{50pt}}

\title{Odd-frequency pairing and time-reversal symmetry breaking for repulsive interactions}

\author{Dimitri Pimenov}\email{dpimenov@umn.edu}
\email{dp589@cornell.edu}
\author{Andrey V.\ Chubukov}
\affiliation{William I. Fine Theoretical Physics Institute, University of Minnesota, Minneapolis, MN 55455, USA}

\begin{abstract}
We study the pairing of fermions by an interaction consisting of a Hubbard repulsion, mimicking a screened Coulomb potential, and a dynamical phonon-mediated attraction. For such interaction, the gap equation allows even- and odd-frequency solutions $\Delta_e$ and $\Delta_o$. We show that odd-frequency pairing does not develop within the Eliashberg approximation due to over-critical pair-breaking from the self-energy. When vertex corrections are included, the pairing interaction gets stronger, and $\Delta_o$ can develop. We argue that even in this case keeping the self-energy is still a must as it cancels out the thermal piece in the gap  equation. We further argue that $\Delta_o$  is not affected by Hubbard repulsion and for strong repulsion is comparable to a reduced $\Delta_e$. The resulting superconducting state is a
superposition $\Delta_e \pm i \Delta_o$, which spontaneously breaks the time-reversal symmetry, despite that the pairing symmetry is an ordinary $s$-wave.
\end{abstract}

\maketitle

\section{Introduction}

When two electrons in a superconductor form a Cooper pair with gap function $\Delta(\r,t)$, they must obey the Pauli principle. This iron fact enables a systematic symmetry classification of
  superconducting order parameters. Interestingly, the Pauli principle can be obeyed even when $\Delta(\r, t)$ is odd under time exchange, $\Delta_o(\r,t) = - \Delta_o(\r,-t)$, Refs. \cite{berezinskii1974new, PhysRevB.45.13125, RevModPhys.91.045005, PhysRevB.45.13125, PhysRevB.79.132502, doi:10.1143/JPSJ.80.054702}. The Fourier transform of such a gap function is an odd function of frequency
  along the Matsubara axis, where $\Delta_o (\k, \omega_m) =-\Delta_o (\k, -\omega_m) $ can be made real  by a proper choice of the phase~\cite{RevModPhys.91.045005}. This implies that at $T=0$, $\Delta_o (\k, 0) =0$. Odd-frequency (OF) superconductivity has a number of profound physical consequences: e.g., it leads to an $s$-wave superconductivity with no gap in the density of states at $T=0$  without magnetic impurities.  OF pairing
  %AC
   has been argued to develop in a two-channel Kondo model~\cite{Kivelson_1992} and
  is particularly favorable in  disordered electron liquids~\cite{Kirkpatrick_1991,zyuzin2019odd} and
  heterostructures \cite{RevModPhys.77.1321, Golubov_2009}, where it was argued to be observed in experiments \cite{DiBernardo2015a, PhysRevX.5.041021}. It  also has a rich interplay with topological effects \cite{doi:10.1143/JPSJ.81.011013}.
   %AC end
   OF pairing can also be induced by an external magnetic field \cite{doi:10.1143/JPSJ.81.033702, PhysRevB.92.054516, PhysRevB.85.224509, APERIS2020168095,tsvelik2019superconductor}.

However, in a regular bulk superconductor at zero field, OF superconductivity remains mostly elusive
 despite being intensively searched for over the last three decades. From a theoretical perspective, there are three obstacles to OF superconductivity.
First, an OF solution $\Delta_o(\omega_m)$ does not emerge at weak coupling
 because the vanishing of $\Delta_o (0)$ implies that there is no enhancement of the pairing susceptibility
  by   the Cooper logarithm. Second,  OF pairing is eliminated by the development of  even-frequency  (EF) superconductivity, which reduces the kernel in the OF pairing channel \cite{RevModPhys.91.045005}.
   Third, even if EF superconductivity does not develop for some reason,  OF pairing is destroyed by pair-breaking effects from fermionic self-energy~\cite{PhysRevB.47.513}.

While the strong-coupling requirement cannot be avoided, we argue there
 are ways to overcome the two other obstacles. In this work we revisit them in the context of phonon-mediated superconductivity, by studying a model of spin-$1/2$ fermions with an effective
 dynamical interaction
\begin{align}
\label{Vdef_a}
 V(\Omega_m)  \propto \lambda \left(f - \frac{\Omega_1^2}{\Omega_1^2 +\Omega^2_m} \right).
\end{align}
This model captures the competition between Hubbard
 repulsion, parametrized by $f$, and phonon-mediated attraction~\cite{PhysRev.125.1263, PhysRevB.28.5100,PhysRevB.100.064513,PhysRevB.94.224515,PhysRevB.96.235107, PhysRevB.98.104505, doi:10.1143/JPSJ.80.044711, PhysRevB.105.064518, pimenov2021quantum}.

 Phonon-mediated OF superconductivity has been analyzed before, most notably by A. Balatsky and co-workers
 (see \cite{RevModPhys.91.045005} and references therein).  They, however, focused on spin-singlet
  $\Delta_o (\k, \omega)$, which is odd in both $\k$ and $\omega$.  We follow the original proposal by
  Berezinskii~\cite{berezinskii1974new} and consider OF superconductivity in the spin-triplet, spatially even channel
  $\Delta_o (\k, \omega) = \Delta_o (-\k, \omega)$, $\Delta_o (\k, \omega) = -\Delta_o (\k, -\omega)$.
    For such superconductivity the momentum dependence of the interaction is not relevant, and one can approximate the electron-phonon interaction by the interaction with an Einstein phonon, as in Eq. (\ref{Vdef_a}).

     The model of Eq.\ (\ref{Vdef_a}) has been analyzed in Ref. \cite{doi:10.1143/JPSJ.80.044711} for particular $\lambda$ and $\Lambda$ and without including fermionic self-energy into consideration. We extend the analysis of \cite{doi:10.1143/JPSJ.80.044711} to arbitrary parameters and also analyze how the results change  when the self-energy is included. A numerical analysis of the effects of fermionic self-energy and vertex corrections for phonon-mediated OF superconductivity  has been recently  performed in Ref.\ \cite{PhysRevB.104.174518}. Where applicable, our results are in agreement with this work.

 We first analyze OF superconductivity in the model of
  Eq. (\ref{Vdef_a}) on its own, assuming the EF superconductivity is not present.
  We prove a compact and fairly general theorem that OF superconductivity cannot develop within the
  canonical Eliashberg approximation, in which the interaction that gives rise to the pairing is exactly the same one that determines the fermionic self-energy. More specifically, we show that    thermal contributions from the static interaction to the pairing vertex $\Phi_o (\omega_m)$ and the self-energy $\Sigma (\omega_m)$ are    exactly the same     and cancel out in the gap equation for $\Delta_o (\omega_m) = \Phi_o(\omega_m) /(1 + \Sigma (\omega_m)/\omega_m)$, but the non-thermal piece
      is stronger for the self-energy,  and  this does not allow OF superconductivity to develop.
    We then go beyond this  approximation, include vertex corrections, and show that the dressed interaction in the particle-particle channel becomes different from the one in the particle-hole channel. We show that  in our model the pairing interaction gets relatively stronger and, as a result,  OF superconductivity  does develop at strong enough coupling. This is in agreement with Ref.~\cite{PhysRevB.104.174518}, where OF  superconductivity has been obtained numerically in ``vertex-corrected" Eliashberg theory~\cite{PhysRevB.102.024503}.

We compute the onset temperature for OF pairing, $T_c$, and show that a non-zero OF condensate develops below $T_c$.  We argue, however, that the self-energy cannot be neglected entirely as thermal contributions to  the pairing vertex and the self-energy still cancel out even when vertex corrections are present.

We next analyze the interplay between OF and EF superconductivity.   It has been shown previously~\cite{doi:10.1143/JPSJ.80.044711, PhysRevB.100.180502}  that static Hubbard repulsion suppresses EF superconductivity, but cancels out in  the gap equation for OF pairing. Taken at a face value, this would imply that OF superconductivity  wins at sufficiently strong Hubbard repulsion.
 We show that the situation is more complex: While Hubbard repulsion is always detrimental to EF pairing, it does not eliminate it completely at strong coupling, which we need for OF pairing (by strong coupling we mean large overall coupling constant $\lambda$ in Eq.\ (\ref{Vdef_a})). The argument is that
   the EF gap function $\Delta_e (\omega_m)$
 changes sign between small and large frequencies, and for large enough $\lambda$
  the system chooses the position of the gap change
  to almost completely eliminate the effect of the Hubbard $f$.  The outcome is that EF superconductivity survives even when $f$ tends to infinity, and the corresponding $T_c$ for EF pairing does not depend on $f$.   We argue that in this situation the onset temperature for EF pairing is still higher, but  the one for OF pairing becomes comparable when the Debye frequency $\Omega_1$ in (\ref{Vdef_a})  becomes comparable to the Fermi energy, which     acts as the UV cutoff for the model.  This condition can be realized in low-density materials such as SrTiO$_3$ \cite{PhysRevB.94.224515,PhysRevB.98.104505,GASTIASORO2020168107}, Bi \cite{doi:10.1126/science.aaf8227} and Half-heusler compounds \cite{doi:10.1126/sciadv.1500242}.  When the onset temperatures for EF and OF pairing are comparable,  both $\Delta_e$ and $\Delta_o$  become non-zero below a certain $T$.  We show that the superconducting condensation energy is the largest when the phases of the two gap functions differ by $\pm \pi/2$, i.e., the order parameter is  $\Delta_e \pm i \Delta_o$. Such an order spontaneously breaks time-reversal symmetry, even though the paring symmetry is still  an ordinary $s$-wave.

This paper is structured as follows: In Sec.\ \ref{modelsection} we introduce the model and the gap equation for the EF and OF components. In Sec.\ \ref{noselfsec} we momentarily neglect the self-energy  and analyze the appearance of the OF solution for the gap once the coupling exceeds a certain threshold.  Furthermore, we discuss the special role of the thermal term in the gap equation. In Sec.~\ref{self-energy sec} we include the self-energy, analyze the set of coupled Eliashberg  equations for the pairing vertex and the self-energy and show that the thermal term gets cancelled in the gap equation, and that there is no solution for a non-zero OF gap function. We then go beyond the Eliashberg approximation, evaluate vertex corrections at $T = 0$ and show that they increase the interaction in the particle-particle channel compared to that in the particle-hole channel and make OF superconductivity possible at strong enough coupling.  In Sec.\ \ref{cancellation sec} we analyze vertex corrections at a finite $T >0$ and show that the thermal term in the gap equation still cancels out. In Sec.~\ref{interplaysec} we discuss the interplay between EF and OF superconductivity: First, in  Sec.~\ref{evensec}, we analyze the suppression of the EF solution by a static repulsion and show that once the coupling $\lambda$  exceeds a critical value, EF superconductivity exists for arbitrary strong Hubbard repulsion $f$.  Then, in  Sec.~\ref{Tccompsec}, we compare the couplings and critical temperatures for EF and OF solutions. Finally, in Sec.\ \ref{trsbsec}, we discuss the co-existence of EF and OF gap functions and argue that in such a state time-reversal invariance is spontaneously broken. Conclusion and outlook are presented in Sec.~\ref{concsec}. Technical details are
relegated to the Appendices.

\section{Model and gap equation}
\label{modelsection}

We consider a system of spin-$1/2$ particles that interact via a dynamical interaction \cite{PhysRev.125.1263, PhysRevB.28.5100,PhysRevB.100.064513,PhysRevB.94.224515,PhysRevB.96.235107, PhysRevB.98.104505, PhysRevB.105.064518, pimenov2021quantum}:
\begin{align}
\label{Vdef}
 V(\Omega_m) = \frac{2}{\rho} \times \chi(\Omega_m), \quad \chi(\Omega_m) =  \lambda \left(f - \frac{\Omega_1^2}{\Omega_1^2 +\Omega^2_m} \right) ,
\end{align}
where $\Omega_m$ is a bosonic Matsubara frequency, $\rho$ is the density of states, $\lambda$ a dimensionless coupling constant, $f$ a dimensionless measure of the  Hubbard
 repulsion, and $\Omega_1$ is a typical phonon energy scale, e.g., Debye energy. To distinguish between $T=0$ and a finite $T$   in the calculations on the Matsubara axis,   we will label fermionic and bosonic frequencies as $\omega, \Omega$
   in the $T \rightarrow 0$ limit, and as $\omega_m, \Omega_m$
at a finite $T$. We will measure all frequencies in units of $\Omega_1$ and set  $\Omega_1 \equiv 1$.

The dynamical interaction (\ref{Vdef}) gives rise to perturbations in both the particle-particle and particle-hole channels. Without vertex corrections (the terms that dress $V(\Omega_m)$), it yields a set of two coupled Eliashberg equations for the dynamical pairing vertex $\Phi(\omega)$ and the self-energy $\Sigma(\omega)$. These two equations can be combined
 into a closed-form equation for
the dynamical gap function $\Delta(\omega) = \Phi(\omega)/(1 + \Sigma(\omega)/\omega)$ (see, e.g., Ref.\ \cite{PhysRevB.102.024524} and App.\ \ref{vertexApp}).
We assume that the EF gap function $\Delta_e$ is spin-singlet, and OF $\Delta_o$ is spin-triplet
and do not write spin indices explicitly.  One can easily verify that the gap equation
 has the same form for both $\Delta_e$ and $\Delta_o$. At $T=0$,
\begin{align}
\label{pairingEliash}
\Delta(\omega) = - \int_{-\Lambda}^\Lambda d\omp \chi(\omega - \omp) \times \frac{\Delta(\omp) - \Delta(\omega) \frac{\omp}{\omega}}{\sqrt{(\omp)^2 + |\Delta(\omp)|^2}}\  ,
\end{align}
where the second term in the numerator on the r.h.s. is the contribution coming from the self-energy.
 The   dimensionless
   UV cutoff $\Lambda$ is generally of order of the Fermi energy
  in units of $\Omega_1$. For most metals $\Lambda \gg 1$, but for low-density systems, $\Lambda \simeq 1$.
We will study both $\Lambda \gg 1$  and $\Lambda \simeq 1$. {Since we consider a momentum-independent interaction for simplicity, the resulting gap function $\Delta(\omega)$ has $s$-wave symmetry; more general interaction potentials can also lead to $d$-wave states etc.\ }

The gap equation can be re-expressed by introducing  even and odd components: $\Delta_e(\omega) = \Delta_e(-\omega)$, $\Delta_o(\omega) = -\Delta_o(-\omega)$
~\cite{doi:10.1143/JPSJ.81.033702,PhysRevB.98.104505, RevModPhys.91.045005, doi:10.1143/JPSJ.80.044711}:
\begin{widetext}
\begin{align}
\label{gapsplitting}  &\Delta_{e/o}(\omega) = - \frac{1}{2}
 \int_{-\Lambda}^\Lambda d\omp \frac{1}{\sqrt{(\omp)^2 +
 |\Delta_{e}(\omp)+\Delta_{o}(\omp)|^2}  } \times \left ( \chi_{e/o}(\omega, \omp) \Delta_{e/o}(\omp)  - \chi_o(\omega, \omp) \Delta_{e/o}(\omega)  \frac{\omp}{\omega} \right)  \\ & \chi_{e}(\omega, \omp) = \chi(\omega - \omp) + \chi(\omega + \omp) = 2\lambda f - \frac{\lambda}{1 + (\omega - \omp)^2} -  \frac{\lambda}{1 + (\omega + \omp)^2}    \\ & \chi_o(\omega, \omp) = \chi(\omega - \omp) - \chi(\omega + \omp) = - \frac{4\lambda\times \omega \omp}{\left( 1+ (\omega - \omp)^2 \right) \left( 1+ (\omega + \omp)^2 \right) }  \ .
\end{align}
\end{widetext}
Viewed separately, $\Delta_e (\omega)$ and $\Delta_o(\omega)$  can be made real.
 Observe that
  the Hubbard repulsion $f$ is present in $\chi_e$, but drops out of $\chi_o$ and that the self-energy contribution (the last term on the r.h.s.\ of Eq.~\eqref{gapsplitting}) contains $\chi_o$ for both $\Delta_{e/o}$, i.e., it does not contain $f$.  This last result is a consequence of putting a symmetric cutoff on the fermionic $\omega'$ rather than on a bosonic $\omega'-\omega$, like in canonical Eliashberg theory. If we used the canonical expression,  we  would find that the Eliashberg self-energy  does contain a term which depends on $f$. In the normal state this term  is
\begin{align}
\label{Sigma}
\Sigma(\omega) = - \lambda f\int^{\Lambda + \omega}_{\Lambda - \omega}
 d\omega^\prime
\end{align}
  in the sign convention
  such that $G^{-1} (\k, \omega) = i\left(\omega + \Sigma (\omega)\right) -
 \xi(\k)$, with $\xi(\k)$ the electron dispersion. {For simplicity, we work with a parabolic electron dispersion, $\xi_\k = |\k|^2/2m - \mu$, which can be linearized near the Fermi surface. Eq.\ \eqref{Sigma}} would yield $\Sigma (\omega) = -2\lambda f \omega$ and let to
 the unphysical result that the quasiparticle residue $Z = 1/(1-2\lambda f) >1$. We argue that this is an artefact.  The issue can be traced back to the applicability of the canonical Eliashberg-type
     treatment  of the self-energy for a model  with         frequency-independent
        Hubbard repulsion. We recall that the Eliashberg self-energy is obtained by integrating over $\xi(\k)$ in infinite limits,
  {\it before} integrating over frequency.  This procedure is well justified when the interaction drops
   starting from   frequencies below the cutoff, which is the case for the electron-phonon term, but
   it is not justified for the frequency-independent Hubbard term.
   Indeed, if we  compute the self-energy  to first order in $f$  by integrating over frequency first, we find that it is purely static and    just renormalizes the chemical potential.      The implication is that the
        dynamical $-2\lambda f \omega$  term in the  self-energy is a parasitic one. It can be eliminated by either keeping the cutoff in the bosonic propagator, but adding a ghost counter-term to the Eliashberg self-energy, or by imposing a symmetric cutoff on the integral over fermionic $\omega'$ rather than bosonic $\omega' -\omega$. This is what we did in Eqs.\ (\ref{pairingEliash}) and (\ref{gapsplitting}).  Either way, the Hubbard $f$ term does not contribute to the dynamical self-energy, and the quasiparticle $Z$ remains below $1$.  We verified that for the equation for the pairing vertex and for vertex corrections, which we consider later, the frequency integrals are UV-convergent, and there is no difference between placing a symmetric cutoff on a fermionic or a bosonic  frequency.

\section{Properties of the gap equation for odd-frequency pairing}

 In this section we solve the gap equation for $\Delta_o$, assuming that $\Delta_e$ is absent.

\subsection{
 Without fermionic self-energy}
\label{noselfsec}

\begin{figure}
\centering
\includegraphics[width=\columnwidth]{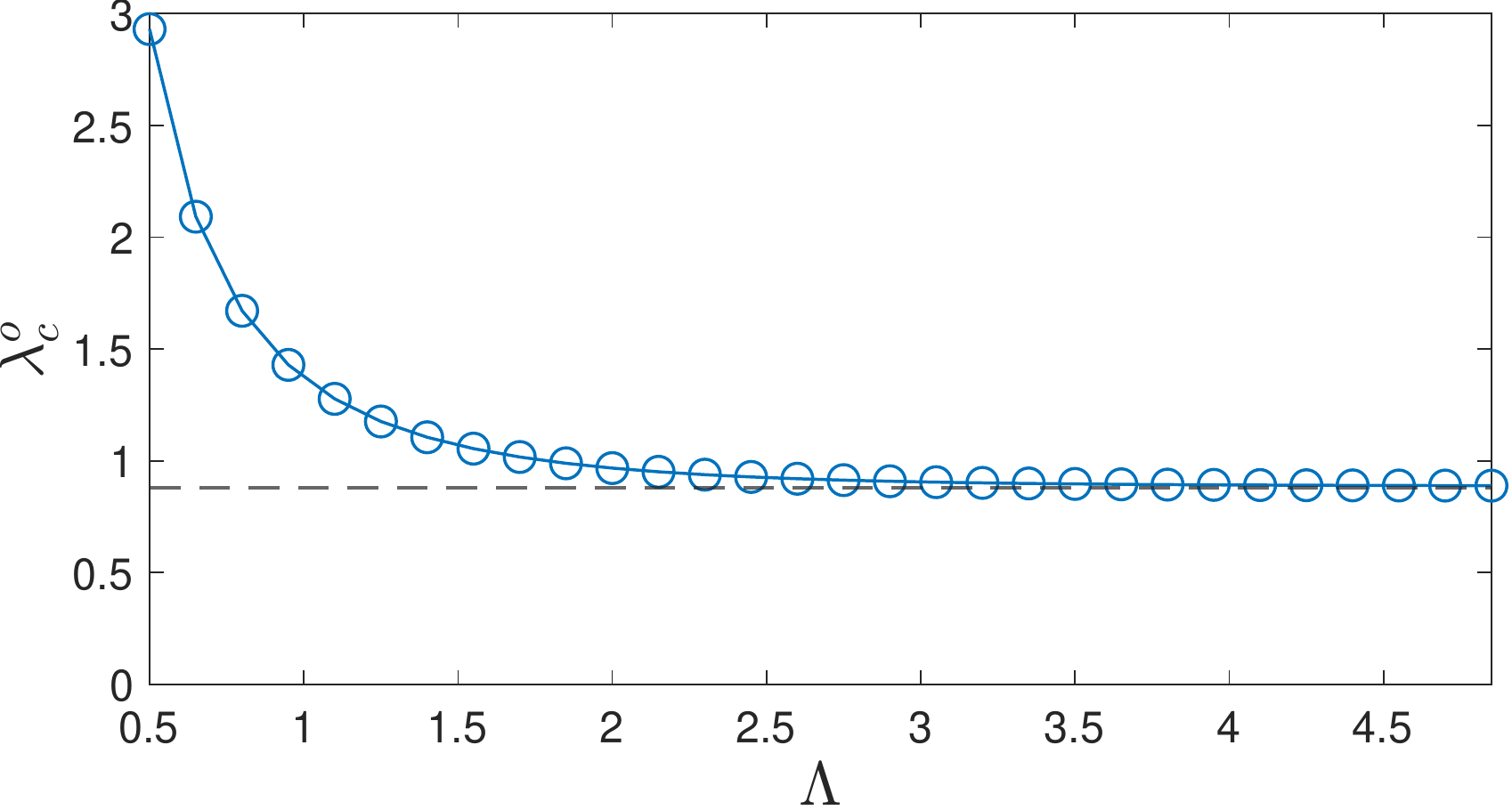}
\caption{Critical value of the coupling, $\lambda_c^o$, as a function of the cutoff $\Lambda$. Odd-frequency superconductivity at $T=0$ develops when $\lambda > \lambda_c^o$.}
\label{lambdaofig1}
\end{figure}

It is instructive to first solve the gap equation for $\Delta_o (\omega)$  without fermionic self-energy.
We remind that the self-energy accounts for the second term in the numerator on the r.h.s.\ of Eq.\ \eqref{gapsplitting}. Neglecting this term, we obtain the truncated gap equation
 \begin{align}
  \label{nn_1}
  &\Delta_{o}(\omega) = - 4 \lambda \omega
 \int_{0}^\Lambda d\omp  \frac{\omp \Delta_{o}(\omp)}{\sqrt{(\omp)^2 + \Delta_{o}^2(\omp)} }  \\ & \notag \times \frac{1}{\left( 1+ (\omega - \omp)^2 \right) \left( 1+ (\omega + \omp)^2 \right) } \ .
  \end{align}  This equation can be solved numerically by  iteration.
  At small $\lambda$, there is no non-zero solution for $\Delta_o (\omega)$ because the pairing kernel is not logarithmically singular.  However at large enough $\lambda$, exceeding the critical one, $\lambda_c^o$, which depends on $\Lambda$,   the solution does exist.  We show $\lambda_{c}^o$ as a function of $\Lambda$ in Fig.~\ref{lambdaofig1}.
   The critical coupling decreases with increasing $\Lambda$ and saturates at $\lambda_c^o \simeq 0.88$ at $\Lambda \to \infty$.

\begin{figure}
\centering
\includegraphics[width=\columnwidth]{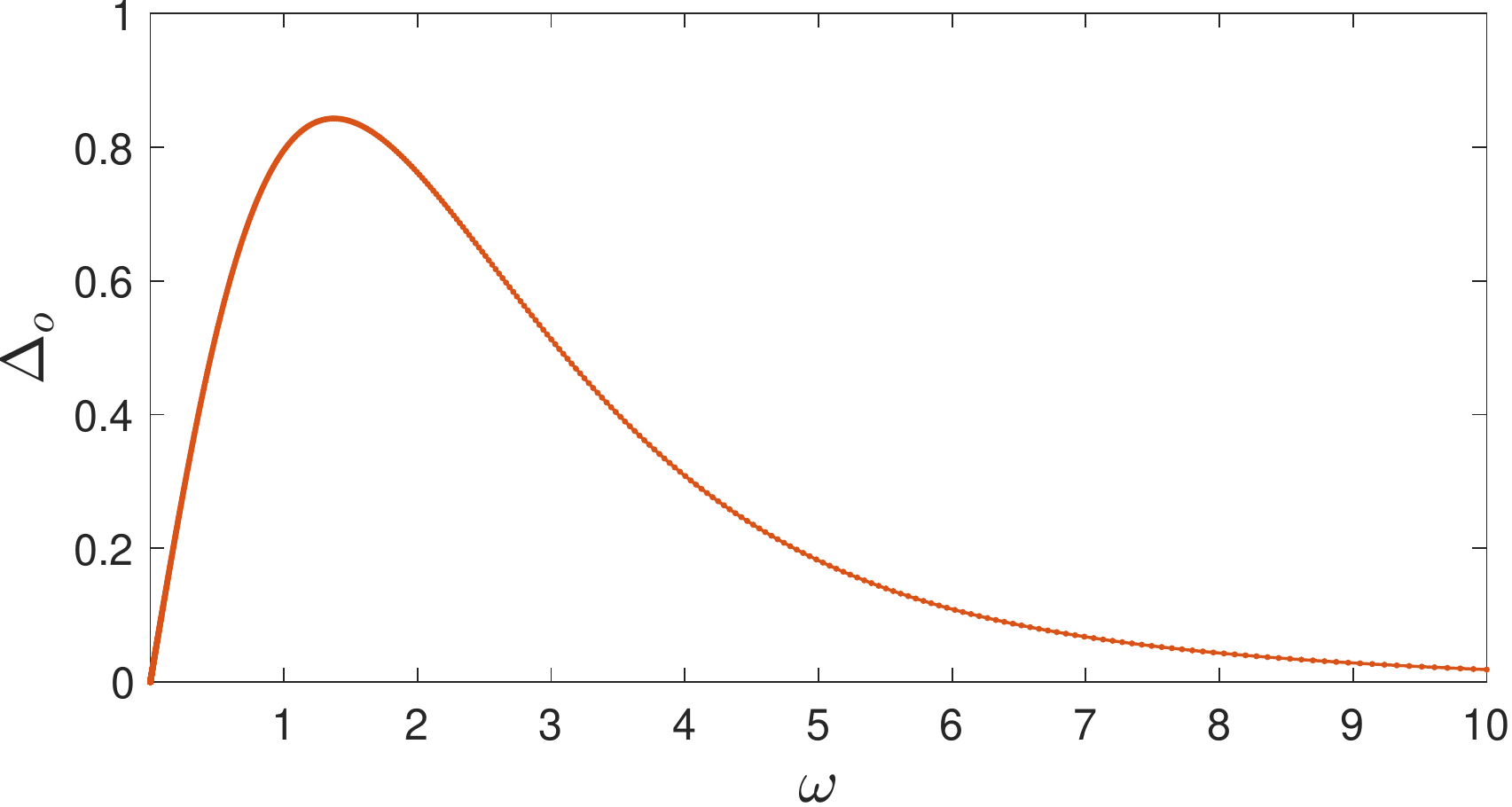}
\caption{
 Odd-frequency  gap function  at $T = 0$ for $\lambda = 1.1, \Lambda = 10$, obtained by solving  the gap equation  without fermionic self-energy.
 }
\label{OFiter}
\end{figure}

In Fig.~\ref{OFiter} we show $\Delta_o$  for $\Lambda=10$ and  representative $\lambda =1.1 > \lambda_c^o$.
     We see that $\Delta_o (\omega)$ scales as $\omega$ at small frequencies, passes through a maximum at a higher $\omega$, and at even higher $\omega$ decreases as $1/\omega^3$.  This last behavior can be obtained analytically by extracting $\omega$ from the kernel on the r.h.s.\ of (\ref{nn_1}) and verifying a posteriori that the remaining integral over $\omega'$ converges.

\begin{figure}
\centering
\includegraphics[width=\columnwidth]{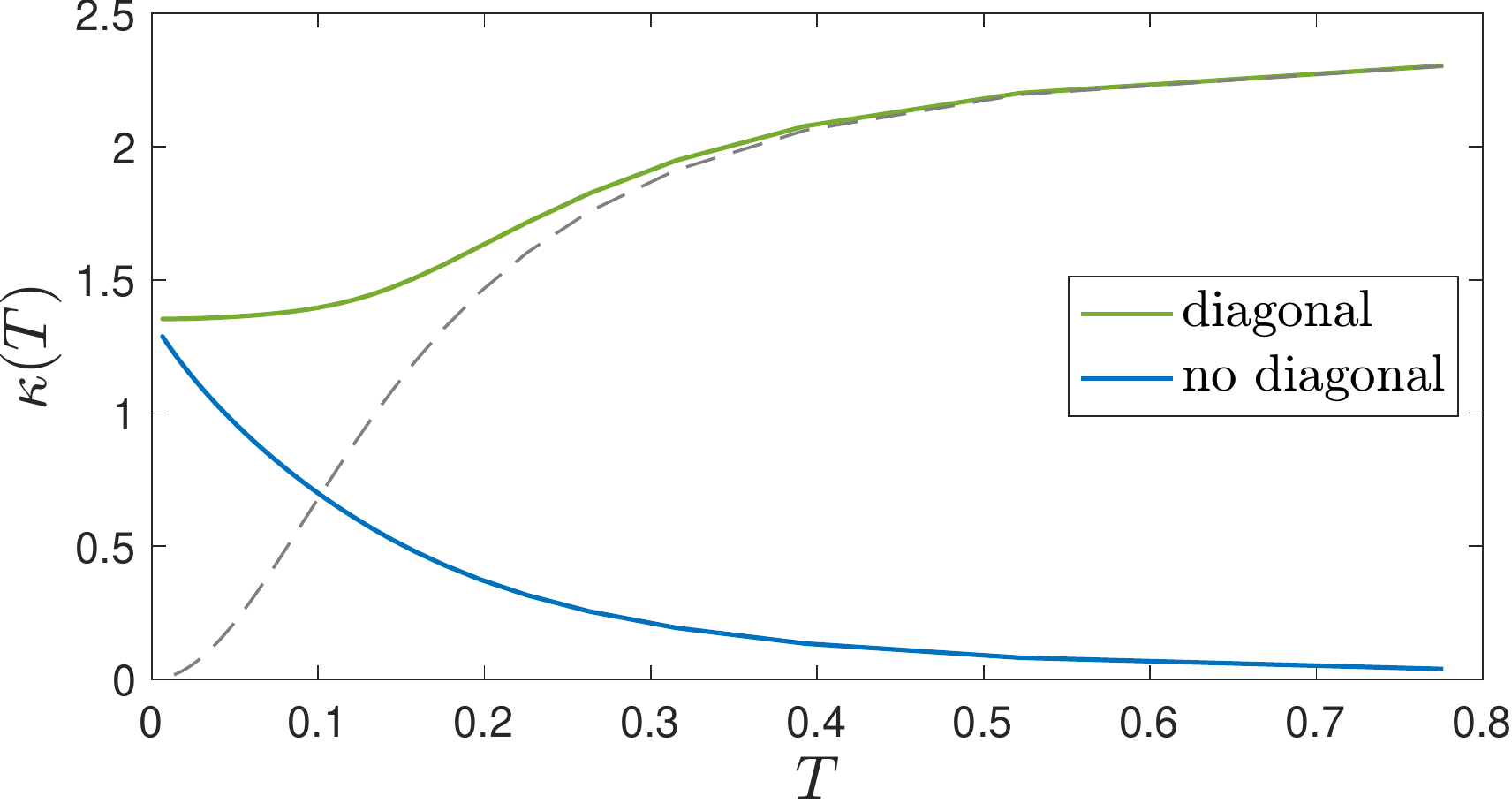}
\caption{Scaling of the largest eigenvalue $\kappa(T)$ of the gap equation kernel $K$,  in the OF case for parameters $\Lambda = 100,  \lambda = 1.2$. The green line shows $\kappa(T)$ when the diagonal elements of $K$ are included, and the dashed gray line corresponds to the eigenvalue estimate of Eq.~\eqref{EVestimate}. The blue line shows $\kappa(T)$ when the diagonal elements are excluded. As expected, the green and blue lines approach the same limit as $T \rightarrow 0$, because the weight of the diagonal entries vanishes in this limit.}
\label{evscalingfig}
\end{figure}

We next move to finite $T$. To obtain the critical temperature $T_c^o$ for OF pairing,  it is convenient to treat the linearized gap equation as a matrix problem. A straightforward discretization of Eq.\ \eqref{nn_1} leads to a matrix equation
\begin{align}
\label{Deltaofirst}
\Delta_o(\omega_m) = \sum_{\omega'_m > 0} K(\omega_m,\omega'_m)\Delta_o(\omega'_m),
\end{align}
where $\omega_m = (2m+1)\pi T$ are (positive) Matsubara frequencies and
 $K$  is the matrix kernel
\begin{align}
\label{Kmatrix}
K(\omega_m,\omega'_m) = T\times  \frac{8\pi \lambda  \omega_m}{(1+ (\omega'_m - \om)^2)(1+(\omega'_m + \om)^2)} \ .
\end{align}
 At the critical temperature $T_c^o$, the largest eigenvalue $\kappa(T)$ of the matrix $K$ is equal to 1.
 It would be natural to expect that $\kappa(T)$ is a decreasing function of $T$, such that $\kappa(T) <1$ at $T > T_c^o$ and  $\kappa(T) >1$ at $T <T_c^o$. However, the numerical analysis yields a different result: $\kappa(T)$ increases with $T$ (the green line in Fig. \ref{evscalingfig}). This leads to a quite exotic  behavior: for $\lambda > \lambda_c^o$, OF superconductivity exists at all $T$, and for $\lambda < \lambda_c^o$, it emerges at some finite $T_c^o$ and exists at larger temperatures. For the model of Eq.\ (\ref{Vdef}) this behavior was first obtained in Refs.~\cite{doi:10.1143/JPSJ.80.044711, doi:10.1143/JPSJ.72.2914}. A similar behavior for spin-singlet superconductivity with gap function odd in both $\k$ and $\omega$ was obtained in the pioneering work of Ref. \cite{PhysRevB.45.13125}.  It was argued in~\cite{doi:10.1143/JPSJ.80.044711} that $\kappa(T)$ is non-monotonic and eventually drops at high enough $T$.  This gives rise to a finite $T_c^o$ for $\lambda > \lambda_c^o$, but to non-monotonic temperature variation of $\Delta_o$ below this temperature, and to reentrant OF superconductivity at $\lambda < \lambda_c^o$, which exists in the window $T_c^o < T < T_{c,2}^o$ (for both end points, $\kappa(T) = 1$).
  This   behavior is reproduced in our model if we impose a UV cutoff on momentum integration, like it was done in~\cite{doi:10.1143/JPSJ.80.044711}.

The exotic behavior of $\kappa(T)$ is easy to understand from Eq.\ \eqref{Kmatrix}:
 because typical $\omega_m$ are of order $T$, off-diagonal elements of $K$ scale a $1/T^2$, while the diagonal elements $K(\omega_m, \omega_m)$, which are thermal contributions from the static interaction $V(0)$, saturate at a finite value at large $T$.  As a result, the largest eigenvalue of $K$ at large enough $T$ is determined  by the largest diagonal element \cite{doi:10.1143/JPSJ.80.044711}, the one at $\omega_m =\omega'_m = \pi T$:
\begin{align}
\label{EVestimate}
T\gg 1: \quad \kappa(T) \rightarrow  K(\pi T, \pi T) = \frac{8\lambda (\pi T)^2}{1 + 4(\pi T)^2}.
\end{align}
We present a numerical check of this behavior in Fig.~\ref{evscalingfig}.
The grey line in this Figure is
$\kappa(T)$ obtained by keeping only the diagonal terms in $K(\omega_{m},\omega'_{m})$
(this is $\kappa (T)$ from  Eq.~\eqref{EVestimate}),
 the green line is the full
$\kappa(T)$. We see that the two expressions coincide at large $T$. We argue below that this exotic behavior is an artifact of neglecting the self-energy.
Indeed, one can see that in the full Eliashberg gap equation \eqref{pairingEliash}, which includes the self-energy,  the thermal contribution with $\omega_m = \omega'_m$ cancels out.
   We show in Sec. \ref{cancellation sec} that this cancellation holds beyond the Eliashberg approximation.
 This cancellation has a drastic effect on where OF superconductivity develops in the $(\lambda, T)$ phase diagram, which we discuss  in detail in the next Section. As a preview, in
 Fig.~\ref{evscalingfig} by a blue line we show the scaling of $\kappa(T)$ for $K(\omega_m, \omega'_m)$ still given by (\ref{Kmatrix}), but with diagonal terms set to zero. We see that $\kappa(T)$ has a conventional behavior: it decreases with increasing $T$. For such $\kappa(T)$, there is no OF superconductivity if $\lambda < \lambda_c^o$, and if $\lambda > \lambda_c^o$, the gap $\Delta_o (\omega_m)$ is non-zero for $T < T_{c}^o$.

\subsection{Role of self-energy at $T = 0$}
\label{self-energy sec}

We  now discuss the self-energy in more detail.
Our first point is that when this term is kept, there is no non-zero solution of the full Eliashberg gap equation (\ref{gapsplitting}) for $\Delta_o (\omega_m)$ at any $T$, including $T=0$.  For interaction with acoustic phonons, this was first observed in Ref. \cite{PhysRevB.47.513}.  In our case of purely dynamical interaction with an Einstein boson, we can prove this explicitly.
 Namely, we argue that if $\Delta_o (\omega_m)$ emerges at some $T_c^o$,  the corresponding
 linearized gap equation at $T$ immediately below $T_{c}^o$ has to satisfy the
 inequality
\begin{align}
\label{theoremmaintext}
\max_n \bigg|\frac{\Delta_o(\omega_n)}{\omega_n}\bigg|  \leq   \frac{  \sum_m  K^T_{n,m} }{ 1 +\sum_m K^T_{n,m} } \times \left(\max_{l} \bigg|\frac{\Delta_o(\omega_l)}{\omega_l}\bigg|\right)  ,
\end{align}
where $K^T_{n,m}$ is the transpose of the OF matrix kernel in Eq.~\eqref{Kmatrix}
 (see App.\ \ref{vertexApp} for details).
 Because all components of $K$ are positive, $\sum_m  K^T_{n,m}/( 1 +\sum_m K^T_{n,m}) <1$, hence a non-zero
  $\Delta_o (\omega_m)$ has to satisfy
the strict inequality
\begin{align}
\label{contra1}
\max_{n} \bigg| \frac{\Delta_o(\omega_n)}{\omega_n} \bigg| < \max_{n} \bigg| \frac{\Delta_o(\omega_n)}{\omega_n} \bigg |\   \quad \text{if} \quad  \Delta_o \neq 0 \ ,
\end{align}
 which is impossible.

When vertex corrections are included, the
 interplay between the attraction in the odd-frequency pairing channel and pair-breaking by the self-energy becomes more nuanced as the interaction in the particle-particle channel, $\chi_{pp}$, and the one in the particle-hole channel,
 $\chi_{ph}$, generally become different. Keeping the two interactions as separate variables in the
gap equation at $T = 0$, we obtain the gap equation in the form
\begin{align}
\label{alphafirst}
&\Delta_o(\omega) = \\ & \notag
 - \int d\omp  \chi_{pp} (\omega -\omp) \times \frac{\Delta_o (\omp) - \Delta_o (\omega) \frac{\omp}{\omega} \times \alpha(\omega -\omp)  }{\sqrt{\Delta^2_o(\omp) + (\omp)^2}} ,
\end{align}
 where
 \beq
 \alpha(\omega -\omp) =  \frac{\chi_{ph}(\omega -\omp)}{\chi_{pp}(\omega -\omp)}\ .
 \eeq
For our purposes, this equation has to be projected to odd-frequency channel.

That vertex corrections make $\chi_{pp}$ and $\chi_{ph}$ non-equivalent can be seen
  already in perturbation theory, by collecting vertex corrections for these two interactions
   to leading order in $\lambda$. We present the diagrams in Fig.~\ref{beyondEmain},  and discuss computational details in Appendix \ref{vertexApp}. We emphasize that the result for the  vertex correction diagram does not depend on whether we impose a symmetric cutoff on an internal fermionic frequency or on a bosonic frequency in the evaluation.

 \begin{figure*}
\centering
\includegraphics[width=\textwidth]{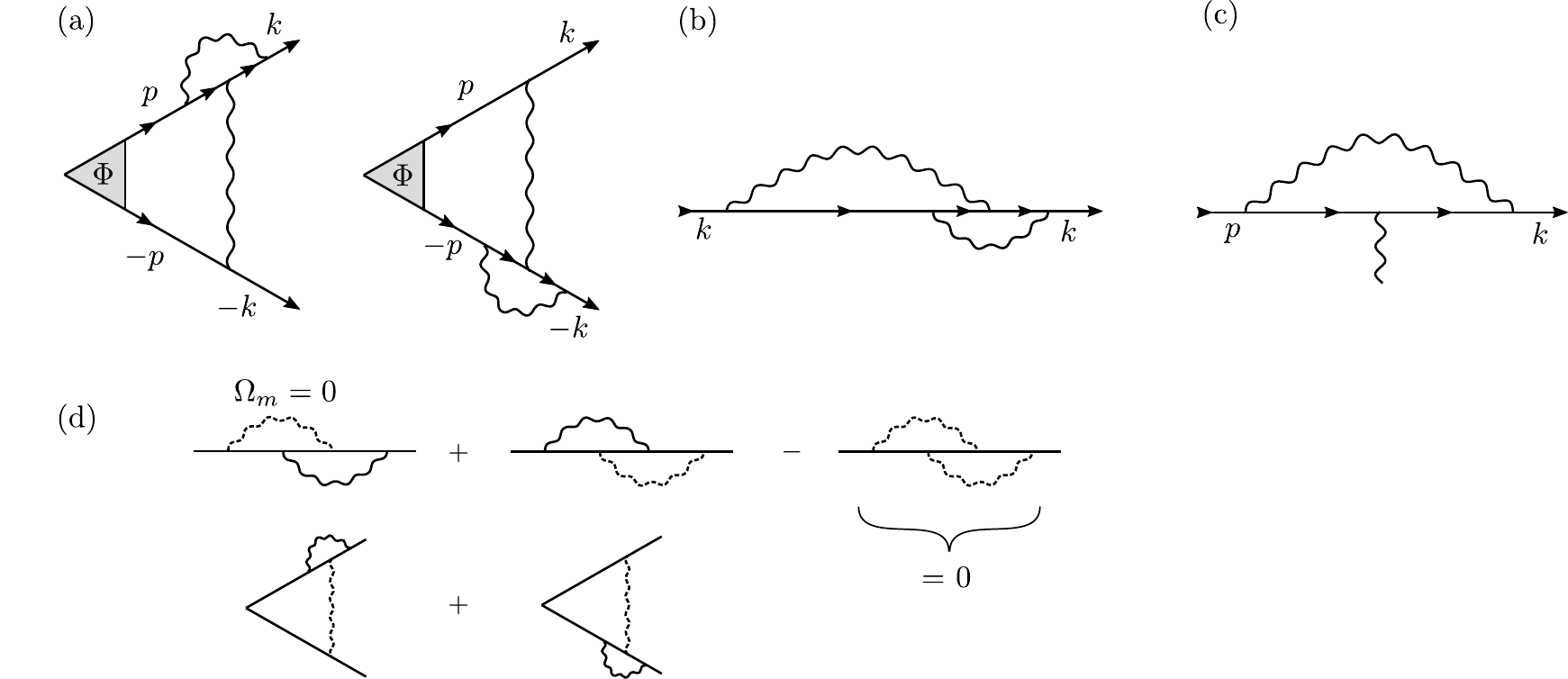}
\caption{(a) Relevant corrections for the pairing vertex $\Phi$. Straight lines represent full Green's functions (including the self-energy), wiggly lines the interaction $V$.  Four-momentum notation is used: $k = (\omega,\k),\ p = (\omega^\prime,\p)$. (b) Relevant correction for the
 self-energy. (c) The vertex correction piece.  (d) Relevant contributions with zero frequency transfer $\Omega_m = 0$, represented by dashed interaction lines. Full interaction lines imply a summation over all frequencies $\Omega_m$. }
\label{beyondEmain}
\end{figure*}

\begin{figure}
\centering
\includegraphics[width=\columnwidth]{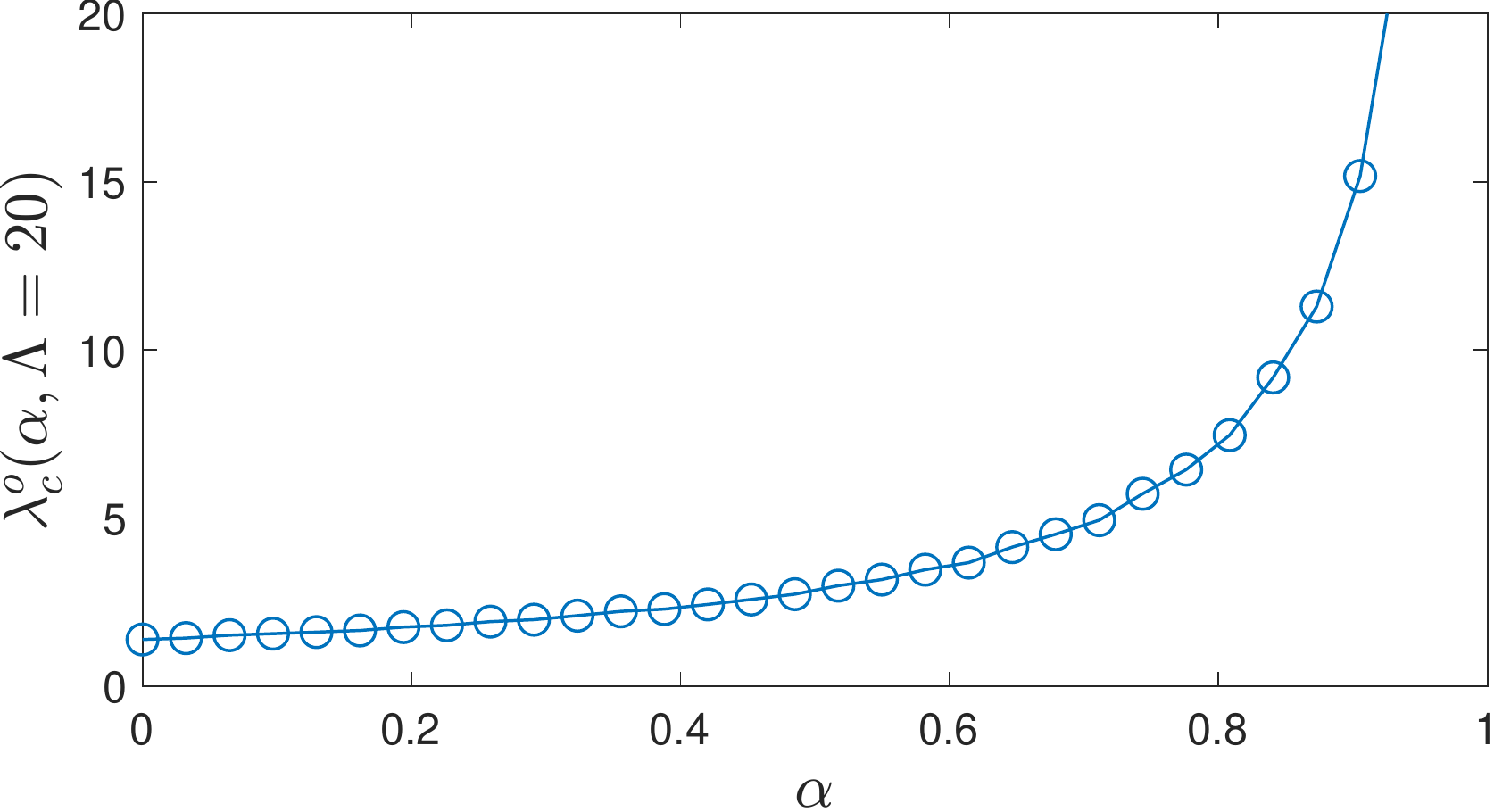}
\caption{Critical value of the coupling, $\lambda_c^o$, for large $\Lambda = 20$ as a function of the self-energy parameter $\alpha$. When $\alpha = 1$, $\lambda_c^o$ diverges and OF superconductivity cannot be realized}
\label{lambdaofig2}
\end{figure}

    The key point is that there are two vertex correction diagrams for the pairing vertex but only one for the self-energy. In both cases, the integration over two fermionic and one bosonic propagator in the vertex correction piece  in Fig.~\ref{beyondEmain}(c) yields $2 \lambda f$. Then under vertex renormalization
\beq
\chi_{pp} \rightarrow \chi_{pp} (1 + 4\lambda f)  , \quad  \chi_{ph} \rightarrow \chi_{ph} (1 + 2\lambda f)
\eeq
 Hence
\beq
 \alpha = \frac{1 + 2\lambda f}{1+ 4\lambda f} < 1 \ .
\eeq
To simplify the analysis, below we treat  $\alpha < 1$  as a phenomenological parameter. In
 Fig.\ \ref{lambdaofig2} we show the behavior of the critical OF coupling $\lambda_c^o$ as a function of $\alpha$, in the limit of large $\Lambda$. At $\alpha =1$ (the original model with no vertex corrections),  $\lambda_c^o = \infty$, which implies that there is no OF pairing for any value of $\lambda$, as we already discussed. However,
  once $\alpha$ becomes smaller than $1$, $\lambda_c^o$ becomes finite, i.e., for strong enough $\lambda$, OF pairing does develop.  At $\alpha \to 0$, $\lambda_c^o$ approaches 0.88, as expected.
The fact that OF pairing develops when vertex corrections are included has also been observed in a recent numerical work \cite{PhysRevB.104.174518}.

 We note in passing that for
   quantum-critical OF pairing by a gapless boson (the limit $\Omega_1 \rightarrow 0$, $\lambda \rightarrow \infty$, $\lambda \Omega^2_1$ tends to a constant), the system is at the boundary towards OF pairing already without vertex corrections.  In this case OF superconductivity emerges already at  infinitesimally small $1-\alpha $ (Ref.~\cite{wu2022odd}).

Also, the authors of Ref.\ \cite{PhysRevB.47.513} argued that the effect of vertex corrections can be modeled by adding a spin-dependent component of the interaction that acts differently in the particle-particle and particle-hole channels. Accordingly, our results can be also modeled by introducing an extra spin-spin component of the interaction.

\subsection{Role of self-energy  at $T > 0$: cancellation of thermal terms in the gap equation}
\label{cancellation sec}

So far we discussed  vertex corrections at $T = 0$. The situation at a finite $T$ is a bit more tricky.
 Namely,  at a finite $T$  we have to distinguish between vertex renormalization of the interaction at a finite frequency transfer $\Omega_m$ and at zero frequency $\Omega_m =0$.  The contributions from the latter to the pairing vertex and the self-energy are associated with thermal fluctuations. Vertex renormalizations to $\chi_{pp} (\Omega_m)$ and $\chi_{ph} (\Omega_m)$ at a finite $\Omega_m$  are essentially the same as at $T=0$, and the interplay between vertex corrections to  $\chi_{pp} (\Omega_m)$ and $\chi_{ph} (\Omega_m)$ is governed by $\alpha <1$.  For interactions with $\Omega_m=0$, computations to the leading (first) order in $\lambda$  yield a different result: there is no factor of 2 difference between vertex corrections to $\chi_{pp} (0)$ and $\chi_{ph} (0)$. To see this, in
 Fig.\ \ref{beyondEmain}(d) we pictorially single out the interactions with $\Omega_m = 0$ by dashed interaction lines. For the pairing vertex at $\Omega_m =0$, there are two different vertex correction diagrams, as before, hence there is an overall factor of $2$. For the self-energy, there is only one diagram, but there are two choices to select which of the two interaction lines carries $\Omega_m =0$ and hence is associated with $\chi_{ph} (0)$. This gives an extra factor of $2$. Then the vertex corrections to $\chi_{pp} (0)$ and $\chi_{ph} (0)$ are the same.  As a result, the thermal piece in the gap equation cancels out even in the presence of vertex corrections.  We conjecture that this holds beyond first order in $\lambda$. We recall that this cancellation eliminated a would-be highly exotic behavior, in which the coupling constant for OF pairing increases with increasing $T$.

To summarize, inclusion of vertex corrections with proper treatment of the thermal terms makes OF superconductivity possible. However, the condition $\lambda > \lambda_c^o = O(1)$ is  required, and $\lambda_c^o$ is large if the vertex corrections are weak. The conditions for OF pairing are
 easier to fulfill in a quantum-critical regime, where the coupling is large.

 \section{Interplay between EF and OF pairing at strong repulsion}
\label{interplaysec}

\subsection{EF superconductivity  and its suppression by static repulsion}
\label{evensec}

\begin{figure}
\centering
\includegraphics[width=\columnwidth]{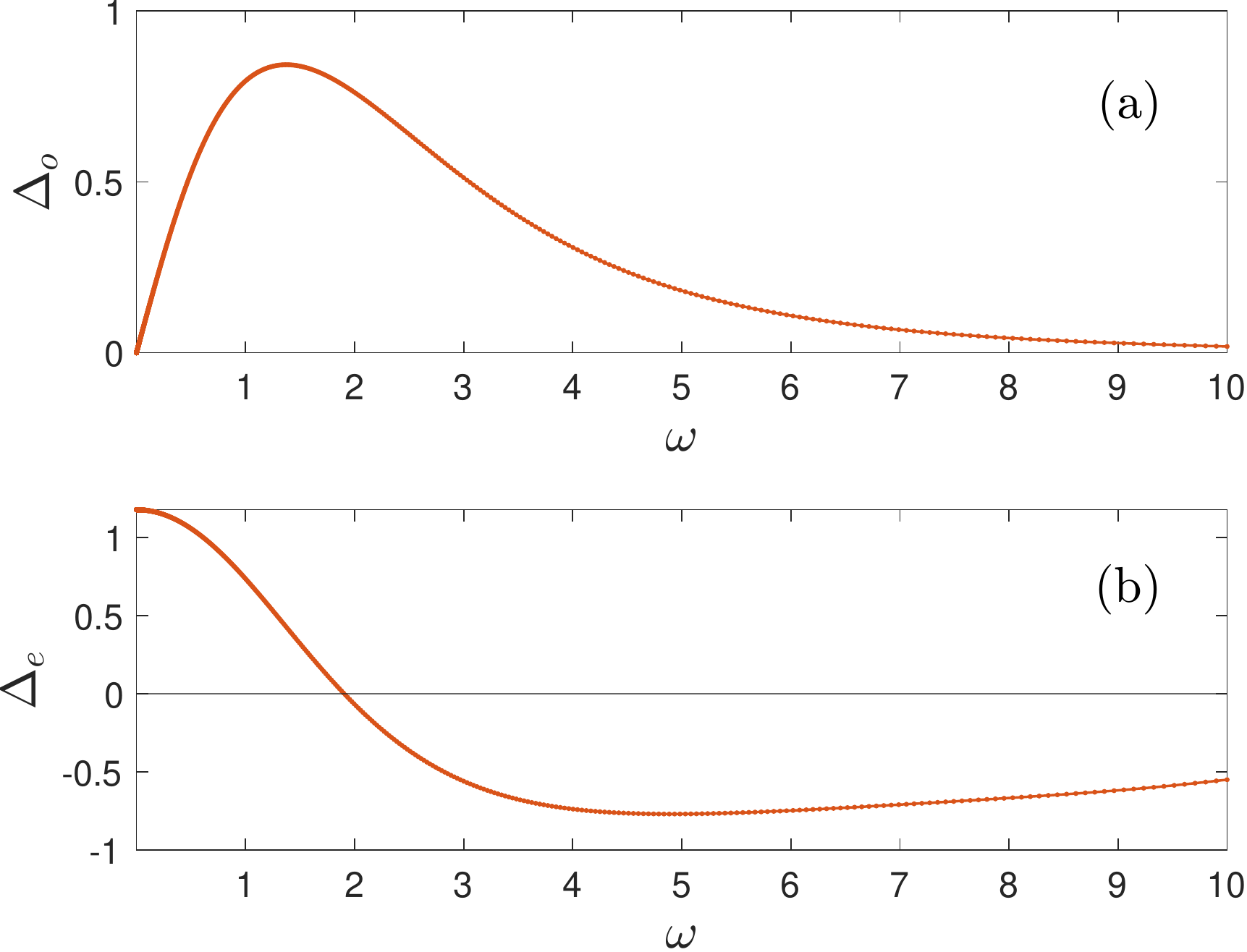}
\caption{
A comparison of  OF (a) and EF (b) gap functions at $T = 0$, obtained by solving
  the gap equation  without the fermionic self-energy.
  We set  $\lambda = 1.1, f = 1.5$ and $\Lambda = 10$.
 }
\label{deltasfig}
\end{figure}

 As we said in the Introduction, a particle-particle interaction of the form of Eq. (\ref{Vdef}) also allows for a conventional superconductivity with even-frequency gap function $\Delta_e (\omega)$.  Below we set  $\Delta_e (\omega)$ to be real
 (we recall that we label by  $\omega$  a continuous Matsubara frequency at $T=0$).
For a non-zero Hubbard repulsion $f$ and $\Lambda \gg 1$,  such $\Delta_e (\omega)$ necessary has a node~\cite{PhysRevB.100.064513,PhysRevB.104.L140501}. A representative $\Delta_e (\omega)$  at $T=0$ is shown in Fig.~\ref{deltasfig}(b).

\begin{figure}
\centering
\includegraphics[width=\columnwidth]{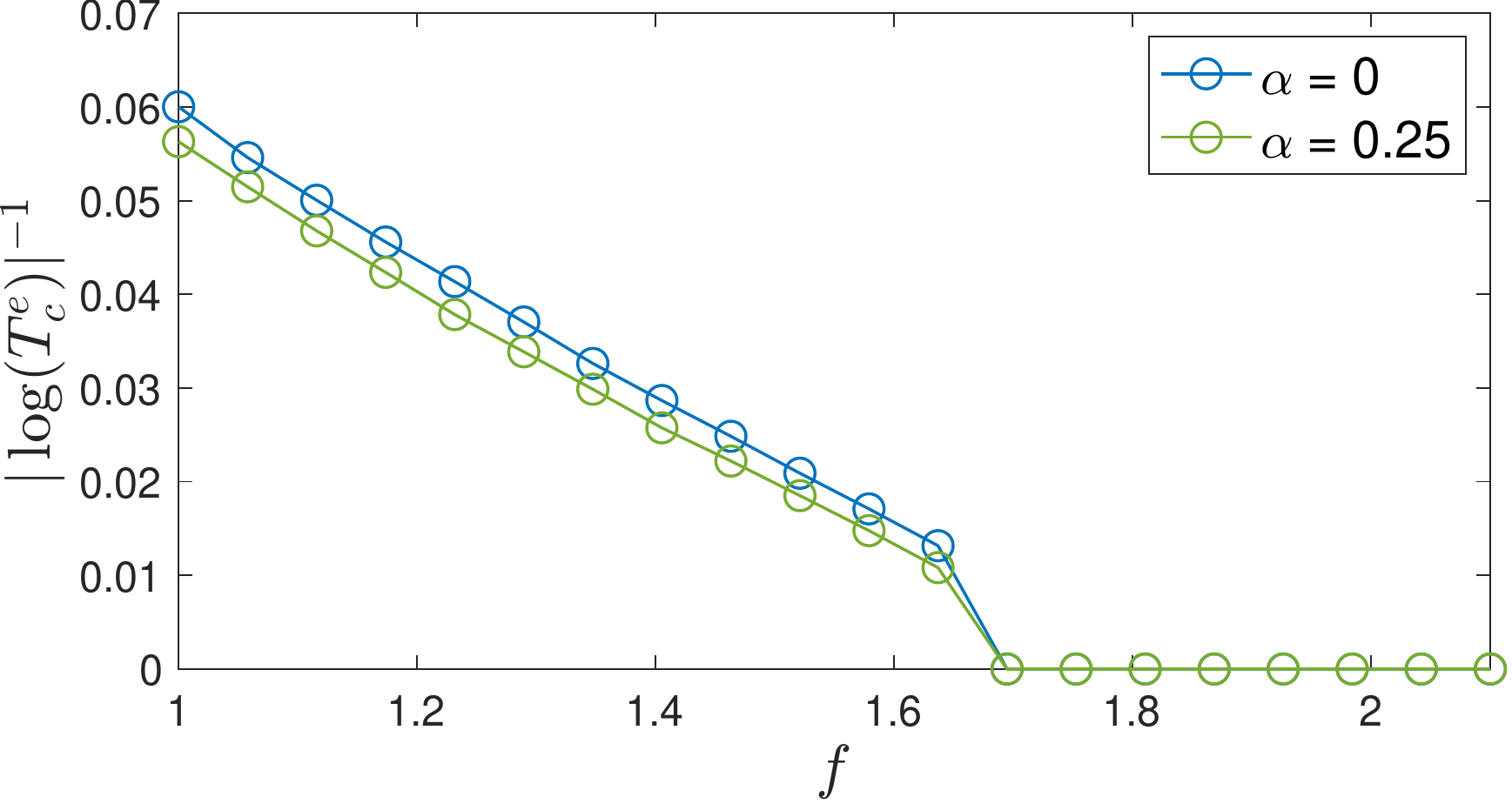}
\caption{$|\log(T_c^e)|^{-1}$ as function of $f$ for two values of $\alpha$,
which measures the strength of pair-breaking effect due to fermionic self-energy
 (at $\alpha =0$ there is no effect from the self-energy). One can see that $|\log(T_c^e)|^{-1}$ scales linearly with $f$, meaning that $T_c^e \sim \exp(1/(f_c - f))$. We set $\Lambda = 10, \lambda = 0.12$. A nonlinear frequency grid was used to reach the required exponentially small temperatures. We have checked that the details of the discretization have no impact on $T_c^e$ near where it vanishes. {Very close to $f_c$, the critical temperatures become too small to be numerically accessible; an extrapolation of $T_c^e$ is shown with dashed lines.} }
\label{Tcsmalllambda}
\end{figure}

For generic $f \leq 1$, $\Delta_e$ is much larger than $\Delta_o$ because of the Cooper logarithm, and for $\lambda < \lambda_c^o$ is the only superconducting solution at $T=0$. When the Hubbard repulsion $f$
 increases, $\Delta_e$ is suppressed. At weak coupling, a phase transition into the normal state occurs when $f$ reaches a critical value $f_c$. For $\Lambda \gg 1$ and $\alpha =0$, this critical value is given by \cite{pimenov2021quantum}
\begin{align}
\label{fcanalytical}
f_c \simeq \frac{1}{1 - 2\lambda \log(\Lambda)} + O(\lambda) \quad\quad  \text{for}\quad  \Lambda \gg 1 \ .
\end{align}
An exemplary ($T$-$f$) phase diagram for EF superconductivity is shown in Fig.~\ref{Tcsmalllambda}. It is obtained by solving the gap equation (\ref{alphafirst}) for infinitesimally small $\Delta_e (\omega)$ for two values of $\alpha$. As expected, the critical temperature for EF pairing, $T_c^e$, vanishes at $f > f_c$. As seen in the Figure, an inclusion of a finite self-energy reduces $T_c^e$ at $f < f_c$, but hardly impacts the value of $f_c$ itself. To understand this, we note that the self-energy  in the even-frequency case still contains the odd component of the dynamical interaction, $\chi_o$, see Eq.\ \eqref{gapsplitting}. Near $f=f_c$, the transition temperature $T_c^e$ is determined by fermions with small frequencies due the IR-divergent Cooper logarithm. Because $\chi_o (\omega, \omega')$ vanishes at $\omega = \omega' =0$, it does not affect the critical $f_c$.

At a first glance the vanishing of $T_c^e$ at $f >f_c$ implies that for such $f$ EF superconductivity is not a competitor to OF superconductivity, and to get OF pairing one just needs to find a way to increase $f$. However, the actual situation is more complex. The reason is that, as we found earlier,  OF superconductivity only holds when $\lambda > \lambda_c^o$, and $\lambda_c^o$ is at least $O(1)$, while  Eq.\ (\ref{fcanalytical}) for $f_c$ only holds for small $\lambda$.  Once we increase $\lambda$, we find that there is another critical value
\begin{align}
\label{lambdae}
 \lambda_{c}^e \simeq \frac{1}{2\log(\Lambda)} \quad  \quad \text{for} \quad     \Lambda \gg 1.
\end{align}
at which $f_c$ diverges. We have verified Eq.\ \eqref{lambdae} numerically and show the results in App.\ \ref{lambdaeApp}.
For $\lambda > \lambda_c^e$,  the EF gap function  is non-zero at $T=0$ for all values of $f$.
  The node in the corresponding $\Delta_e (\omega)$  is placed in such a way that
   $f \times \int d\omega^\prime \frac{\Delta_e(\omega^\prime)}{\omega^\prime} \rightarrow \text{const.}$ when $f \rightarrow \infty$.

\subsection{Critical couplings and temperatures for EF and OF superconductivity}
\label{Tccompsec}

We see from Eq.~\eqref{lambdae} that at large $\Lambda$ and $\alpha =0$,
  $\lambda_c^e \sim 1/\log{(\Lambda)}$ while  $\lambda_c^o \simeq 0.88$ is a constant. Then
  $\lambda_c^e < \lambda_c^o$. This is also true at a finite $\alpha$ as $\lambda_c^o$ increases and  $\lambda_c^e$ remains almost the same. In this situation, it is natural to expect that EF superconductivity prevents the development of OF superconductivity, because when $\Delta_e (\omega)$ becomes non-zero, it reduces the strength of the pairing kernel on the OF channel. This is the last of the three obstacles for OF pairing that we listed in the Introduction.

\begin{figure}
\centering
\includegraphics[width=\columnwidth]{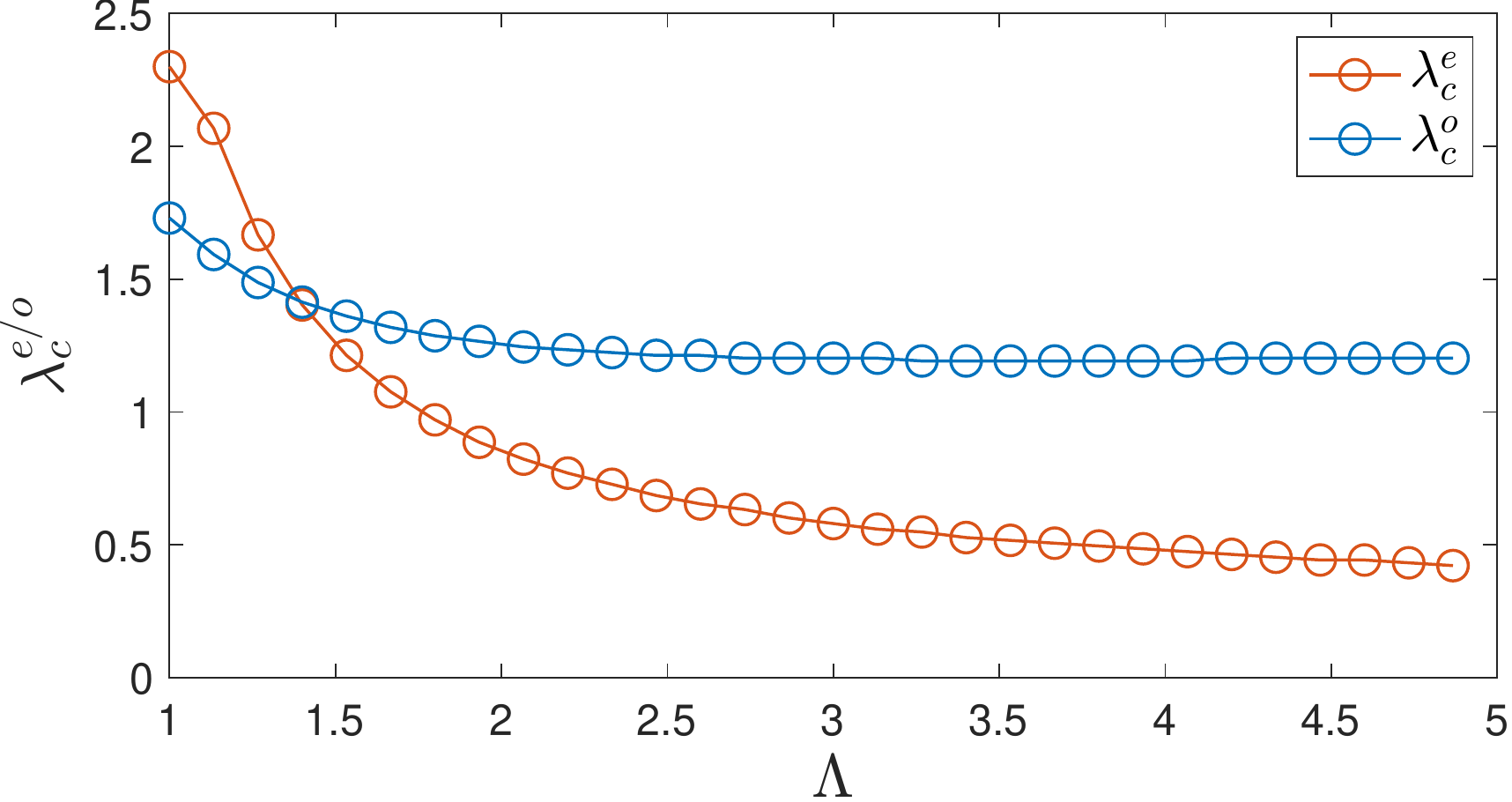}
\caption{Comparison of $\lambda_c^e$ (red) and $\lambda_c^o$ (blue) as a function of $\Lambda$ for $\alpha = 0.2$.}
\label{lambdaefig}
\end{figure}

This obstacle becomes less drastic when $\Lambda = O(1)$, like in low-density materials, e.g.,  SrTiO$_3$ \cite{PhysRevB.94.224515, PhysRevB.98.104505, GASTIASORO2020168107}, Bi \cite{doi:10.1126/science.aaf8227} and Half-heusler compounds \cite{doi:10.1126/sciadv.1500242}.
In this situation $\lambda_c^e$ and $\lambda_c^o$  become comparable, as we show in
 Fig.\ \ref{lambdaefig}.  Correspondingly, $T_c^e$ and $T_c^o$ also become comparable.  There is even a window of $\Lambda$ in Fig.\ \ref{lambdaefig} where OF superconductivity develops first.
  For small $\alpha$, the requirement on $\Lambda$ is even less restrictive. In Fig.\ \ref{deltasfig} we show
$\Delta_o$ and $\Delta_e$, obtained independently by solving the gap equation at $T=0$ at $\alpha=0$, i.e., without the self-energy term, for $\lambda=1.1$, $f =1.5$ and $\Lambda$ as large as 10. We see that even for such large $\Lambda$ the magnitudes of $\Delta_o$ and $\Delta_e$ are comparable.

  These observations suggest  that at $T=0$ both EF and OF gap functions     may be  non-zero. Our next goal is to find   such a mixed state and determine the relative phase factor between the  $U(1)$ order parameters $\Delta_o$ and $\Delta_e$.

\begin{figure}
\centering
\includegraphics[width=\columnwidth]{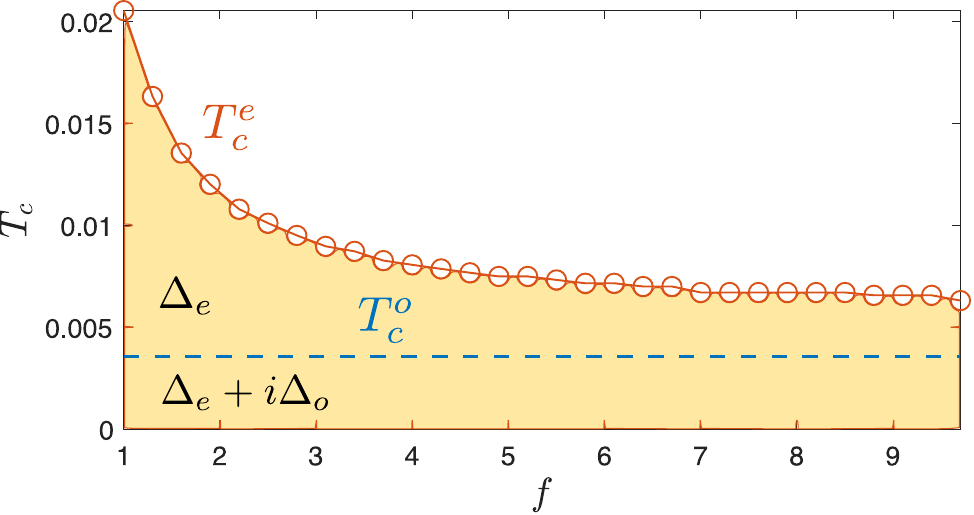}
\caption{Numerical results for  $T_c^e$ (circles) and $T_c^o$ (dashed horizontal line) as a function of $f$.
  We set  $\Lambda = 2, \lambda = 1.5$, $\alpha = 0.3$. Yellow shading denotes the superconducting region. The label $\Delta_e + i \Delta_o$ indicates the superconducting state with broken time-reversal symmetry.}
\label{Tclargelambda}
\end{figure}

\section{Spontaneous breaking of time-reversal symmetry}
\label{trsbsec}

We show an exemplary phase diagram in the $(T,f)$ plane in Fig.\ \ref{Tclargelambda} by choosing a $\Lambda$ for which EF superconductivity develops first below $T_c^e$, but $T_c^o$ is close, and while it is reduced by a finite $\Delta_e$, the OF component still develops below a finite $T$.

As we will demonstrate below, the gap function in the mixed state at $T=0$
is of the form
 \begin{align}
\label{firstsuperpos}
\Delta(\omega) = \Delta_e(\omega) \pm i \Delta_o(\omega) \ .
\end{align}
This agrees with Ref.\ \cite{PhysRevB.104.174518}, where a mixed state with spin-singlet EF and spin-triplet OF order parameter with the relative phase $\pm \pi/2$ has been found numerically (for a cuprate-like Fermi surface and $d-$wave spatial symmetry of both gap functions).

A superconducting state with $\Delta (\omega)$ from Eq.\ \eqref{firstsuperpos}
has a special property: it \textit{spontaneously} breaks time-reversal symmetry, despite that separately spin-singlet $\Delta_e$ and spin-triplet $\Delta_o$ are invariant under time reversal. We show in App.~\ref{Ttrafoapp} that time-reversal $\mathcal{T}$ acts on the  gap function along the Matsubara axis
 simply as a complex conjugation:
\begin{align}
\label{timerevmain}
(\mathcal{T} \circ \Delta)(\omega) = \left[\Delta(\omega) \right]^{*}.
\end{align}
 Taken separately,
 $\Delta_e(\omega)$ and $\Delta_o(\omega)$ are time-reversal invariant
 (we recall that $\Delta_o$ is odd under time permutation, but even under time reversal \cite{RevModPhys.91.045005, PhysRevB.96.174509}).
However, $\Delta (\omega)$ from \eqref{firstsuperpos} does not remain invariant under time reversal.

To see that the relative phase between $\Delta_e$ and $\Delta_o$ is $\pm \pi/2$,  we
 consider the gap equation at $T=0$ without self-energy correction and assume that the gap function is
\begin{align}
\label{mixedansatz}
\Delta(\omega) = \Delta_e(\omega) + \exp(i\phi) \Delta_o(\omega) ,
\end{align}
 and that $\Delta_o$ is smaller than $\Delta_e$. To leading order in $\Delta_o$, the gap equation in the OF channel then takes the form
\begin{align}
&\exp(i\phi) \Delta_o(\omega) = \\ \notag & 2\int_0^\Lambda d\omega^\prime \frac{\chi_o(\omega - \omega^\prime) \Delta_o(\omega^\prime) }{\sqrt{(\omega^\prime)^2 + \Delta_e^2(\omega^\prime)}} \\ \notag &\times \left( \exp(i\phi) - \frac{\cos(\phi) \Delta_e^2(\omega^\prime) }{ (\omega^\prime)^2  + \Delta^2_e(\omega^\prime)}\right) + O(\Delta_o^2(\omega^\prime) )
\end{align}
One can easily verify that $\phi$ can be either zero or $\pm \pi/2$. For $\phi = 0$, the expression in parentheses reduces to $1 - \Delta_e^2/((\omega^\prime)^2 + \Delta_e^2)$, which reduces $\chi_o$.
For $\phi = \pm \pi/2$, the expression in parentheses becomes $\pm i$, in which case there is no suppression.
We conclude therefore that the mixed state with $\phi = \pm \pi/2$ is indeed preferential.
 We note   in passing that the state $\Delta = \Delta_e \pm i \Delta_o$ is also realized when the time-reversal symmetry is broken explicitly by applying a magnetic field, as shown in Ref.\ \cite{doi:10.1143/JPSJ.81.033702}.

A spontaneous breaking of time-reversal symmetry can be detected experimentally via muon spin relaxation or Kerr rotation \cite{Ghosh_2020} and such states have been intensively discussed in recent years, but chiefly for non-s-wave spatial symmetry, or for multi-band s-wave superconductors~\cite{PhysRevB.87.144511}.  In our case a superconducting state with broken time-reversal symmetry emerges in a one-band $s$-wave superconductor.

\section{Conclusion and outlook}
\label{concsec}
In this work we considered  OF superconductivity in a model of fermions with an interaction potential which contains a static  Hubbard repulsion and a dynamical phonon-mediated attraction. We critically reexamined the three foes which usually prevent OF superconductivity: the necessity for strong coupling, the self-energy effect and the suppression by EF superconductivity. We have found that the strong coupling requirement cannot be avoided, but there are ways to overcome the other two obstacles.   The self-energy does prevent OF superconductivity in the Eliashberg approximation, if the same interaction determines the pairing and the  fermionic self-energy. We    argued that vertex corrections change this balance and make OF pairing possible. At the same time, the self-energy cannot be simply neglected as with and without vertex corrections it  leads to cancellation of the thermal terms in the gap equation. Consequently, we find an OF state, which is stable below $T_c^o$ down to zero temperature, and not the reentrant behavior observed in previous works.

The suppression of OF pairing by pre-existing  EF superconductivity remains a problem when the Fermi energy is much larger than the typical phonon energy scale. However, when these scales become comparable, the critical temperatures for EF and OF orders  are  comparable, and OF superconducting order can  co-exist with EF superconductivity. We have shown that a mixed state        with the gap function       $\Delta_e(\omega) \pm i \Delta_o(\omega)$ can be realized in this case. This state spontaneously breaks the time-reversal invariance.

{It has been argued that  induced  OF superconducting state may exhibit a  paramagnetic Meissner effect
     (see Ref. \cite{Cox_98} and references therein).  However, as shown in Refs.\ \cite{PhysRevB.60.3485, PhysRevB.79.132502, doi:10.1143/JPSJ.80.054702} and also discussed  in Ref.\ \cite{RevModPhys.91.045005}, for       spontaneous OF superconductivity in the bulk, induced by a  retarded interaction, the Meissner effect is diamagnetic. This can be seen explicitly by computing the superfluid density $n_s$ (see Appendix \ref{Meissnerapp}), which is manifestly positive. While this result has been questioned in the literature due to possible issues with spontaneous $U(1)$ symmetry breaking \cite{PhysRevB.91.144514}, in our understanding all ambiguities can be avoided by consistently working in the functional-integral formalism (and avoiding a Hamiltonian description). The fact that a diamagnetic Meissner effect is 
"conventional" can also be seen by reformulating the OF theory in terms of $D(\omega) = \Delta_o(\omega)/\omega$, which is an even function of frequency, like $\Delta_e (\omega)$ for EF superconductivity. 
Using the description of OF superconductivity in terms of $D(\omega)$,  one straightforwardly obtains 
 conventional  electromagnetic response of the superconducting state.
A \textit{paramagnetic} Meissner effect can develop for induced OF superconductivity  at a boundary of a system, but this is a different setup from the one considered in this work.
}

 Our analysis of the OF state was performed by studying gap functions on the Matsubara axis. On the other hand,  the measurable physical properties of the system are determined by the gap function on the real axis. Since the OF gap    vanishes at $\omega =0$,    the  density of states of an OF superconductor    is qualitatively similar to that of an EF gapless   superconductor in the presence of magnetic impurities \cite{RevModPhys.78.373}. However, we expect crucial differences in, say,    the phase winding of the gap function    and in the behavior of low-energy collective modes, which could be fruitful objects for future studies.

 {Note added: shortly after this paper was posted, a work appeared \cite{langmannOF} which contains a more general version of the ``No-go-theorem" for OF superconductivity within the Eliashberg approximation
 due to the self-energy effects .}

\section{Acknowledgement}
We thank Shang-Shun Zhang, Prachi Sharma and Zhentao Wang for useful discussions.
The work by A.V.C  was supported by the NSF DMR-1834856.
A.V.C acknowledges the hospitality of KITP at UCSB, where part of the
work has been conducted. The research at KITP is supported by the
National Science Foundation under Grant No. NSF PHY-1748958.

\appendix

\section{Proof that $\Delta_o$ vanishes within the Eliashberg approximation}
\label{theoremApp}

We introduce a function $D(\omega) = \Delta_o(\omega)/\omega$. In terms of $D$, the linearized gap equation \eqref{pairingEliash} at finite $T$ reads
\begin{align}
D(\omega_n) =   \sum_{m} K^T_{n,m} \left[D(\omega_m) - D(\omega_n) \right] ,
\end{align}
where $K^T_{n,m} = K(\omega_m, \omega_n)$ is the transpose of the OF kernel defined in Eq.~\eqref{Kmatrix}.

By bringing $D(\omega_n)$ to the left side, we get
\begin{align}
\label{Rdiscrete}
D(\omega_n) = \sum_m \frac{ K^T_{n,m} }{ 1 +  \sum_m K^T_{n,m} } \times D(\omega_m) \ .
\end{align}
Now we assume that $D(\omega_n)$ is finite for all $n$ and look at $\max_n |D(\omega_n)|$:
\begin{align} \notag
&\max_n |D(\omega_n)|  = \\ & \notag \max_n \big |\frac{  \sum_m  K^T_{n,m} }{ 1 +\sum_m K^T_{n,m} } \times D(\omega_m) \big | \leq \\  \notag & \max_n  \bigg| \frac{   \sum_m K^T_{n,m} }{ 1 +\sum_m K^T_{n,m} }   \bigg  | \times \big| D(\omega_m) \big | \leq \\   & \max_n  \bigg| \frac{  \sum_m  K^T_{n,m} }{ 1 +\sum_m K^T_{n,m} }   \bigg  | \times \max_{l} \big|D(\omega_l) \big | \label{theorem1}\  ,
\end{align}
where the triangle inequality was used. Since $\chi_o$ is attractive, all entries of $K$ are positive. Then we can rewrite Eq.\ \eqref{theorem1} as
\begin{align}
\label{theorem2}
&\max_n |D(\omega_n)|  \leq \notag  \frac{  \sum_m  K^T_{n,m} }{ 1 +\sum_m K^T_{n,m} } \times \max_{l} \big|D(\omega_l) \big | \ . \end{align}
The first factor in the second inequality is smaller than $1$ for any $n$. Assuming that $\max_n |D(\omega_n)| > 0$, we obtain the strict inequalitiy
\begin{align}
\max_n |D(\omega_n)| < \max_n |D(\omega_n)| \ ,
\end{align}
which is a contradiction. Therefore, we must have $D = 0 \Leftrightarrow \Delta_o = 0$.

\section{Vertex corrections}
\label{vertexApp}

\subsection{Evaluation at $T = 0$}

Including the second order diagrams of Fig.~\ref{beyondEmain} in the main text, the linearized Eliashberg equation at $T = 0$ can be written down as
\begin{align} \notag
&\Phi(k) = - \int_p \Phi(p) G(p) G(-p) V(k - p) \times \left[ 1+ 2 \Gamma(k,p) \right]
\\
\label{eliashapp}
& \Sigma(k) = - i\int_p G(p) V(k - p) \times \left [ 1 + \Gamma(k,p) \right] ,
\end{align}
where
\begin{align}
\Gamma(k,p) = - \int_l G(l) G(p + l - k) V(k - l) ,
\end{align}
and we use notations $k = (\omega, \k), p = (\omp, \p), l = (\omt, \lvec)$ and the conventions
\begin{align}
&G(k) = \left(i\omega - \xi(\k) + i\Sigma(k)\right)^{-1} , \\
& \int_k = \int \frac{d\omega d\k}{(2\pi)^d} \ ,
\end{align}
with $\xi(\k)$ the electron dispersion, which can be linearized near the Fermi surface. We will focus on $d = 2$ for concreteness, and comment on the analogous 3d results along the way.

Without the vertex corrections, $\Gamma = 0$, the gap equation \eqref{pairingEliash} directly follows from
\eqref{eliashapp} by rewriting the definition of the gap, $\Delta(\omega) = \Phi(\omega)/(1+\Sigma(\omega)/\omega)$ as
\begin{align}
\Delta(\omega) = \Phi(\omega) - \frac{\Delta(\omega)}{\omega} \Sigma(\omega) \  ,
\end{align}
and evaluating the momentum integrals in \eqref{eliashapp}.

We now evaluate the vertex correction $\Gamma$, using bare Green's functions (no $\Sigma$), and working in the limit $\Lambda \rightarrow \infty$ for simplicity. Shifting the integration variables, $\Gamma$ can be written as
\begin{align}
&\Gamma(k,p) = (-1) \lambda  \int \frac{d\lvec d\omt}{(2\pi)^3}  \frac{2}{\rho} \times \left( f - \frac{1}{1 + (\omt)^2} \right) \\ \notag & \times \frac{1}{i(\omt+ \omp) - \xi_{+}} \frac{1}{i(\omt + \omega) - \xi_{-}} , \\ &  \notag \xi_+ = \xi(\lvec + \frac{1}{2} (\p - \k) ), \quad   \xi_{-} = \xi(\lvec - \frac{1}{2} ( \p - \k) ) \ .
\end{align}
We expand the dispersion as
\begin{align}
&\xi_+ = \xi({\lvec}) + \delta q, \quad \xi_{-} = \xi({\lvec})- \delta q , \\ &   \delta q = \frac{1}{2} v_F  |\q|\cos(\phi), \quad \q  = | \p - \k | , \quad \phi = \measuredangle (\q, \lvec) \notag \ ,
\end{align}
and integrate over $\xi(\lvec)$ in infinite limits.
Such an expansion is legitimate as for $\Omega \equiv \omp - \omega \ll E_F$ the relevant contributions come from small angle scattering where $|\q| \ll |\k|, |\p|$. We call $\Gamma_1$ the part $\propto f$  and the remainder $\Gamma_2$. To compute $\Gamma_1$, we need to perform the frequency integral first, since the integral is not absolutely convergent. The computation is standard; it is the same as for the polarization function, since the part $\sim f$ is short-range. We obtain
\begin{align}
\Gamma_1 = 2 \lambda f \left( 1 - \frac{|\Omega|}{\sqrt{ \Omega^2 + (v_F |\q|)^2} } \right) ,
\end{align}
The second term in $\Gamma_1$ contains $|\q|$. To find the renormalization of the interactions which enter the Eliashberg equations, we can take the $s$-wave part of this term. I.e., we write
\begin{align}
|\q|^2 = |\p - \k|^2= 4 k_F^2 \sin^2(\theta/2), \quad \theta = \measuredangle (\p,\k) ,
\end{align}
where $\k$ and $\p$ are on the Fermi surface. We then average
\begin{align}
\label{A1smallpart}
\int_0^{2\pi} \frac{d\theta}{2\pi} \frac{|\Omega|} {\sqrt{ \Omega^2 + 16 E_F^2 \sin^2(\theta/2)}} \simeq \frac{|\Omega|}{2\pi E_F} \log \left(\frac{ E_F }{ |\Omega|} \right)
\end{align}
In $d = 3$, one obtains a correction $\sim |\Omega|/E_F$ without the logarithm. In the limit $E_F \gg 1$, the dynamical correction is small, in accordance with Migdal's theorem \cite{migdal1958interaction}.

To compute $\Gamma_2$, it is easier to perform the integral over dispersion $\xi_{\l}$ first. This is allowed because the extra frequency-dependence renders the integral absolutely convergent. The result is
\begin{align}
\Gamma_2 = &2\lambda f \text{sign} (\Omega)  \frac{1}{\sqrt{(v_F |\q|)^2 + \Omega^2}} \\ & \times \left[ \arctan \left(\frac{1}{\omega}\right) - \arctan \left( \frac{1}{\omp} \right) \right] \notag \ .
\end{align}
This term depends on $\omega$ and $\omp$ separately. I.e., it depends on both $\omega - \omp$ and $\omega + \omp$. However, after taking the $s$-wave part it will scale as $|\Omega|/E_F$ similar to \eqref{A1smallpart}.

Collecting the results, the vertex correction reads
\begin{align}
\Gamma = \Gamma_1 + \Gamma_2 = 2\lambda f + |\omega - \omp| \times O\left(\frac{1}{E_F}\right) ,
\end{align}
The static part reads $2\lambda f$, as stated in the main text.

\subsection{Additional diagrams}

\begin{figure}
\centering
\includegraphics[width=\columnwidth]{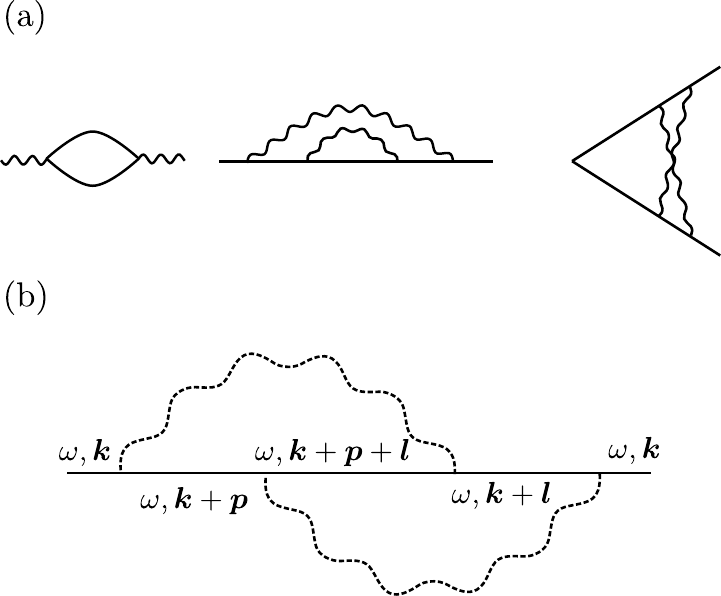}
\caption{(a) Additional second order diagrams not considered in the main text. (b) Self-energy contribution where both interaction lines are thermal. }
\label{beyondEexpra}
\end{figure}

In Fig.\ \ref{beyondEexpra}(a), we show additional second order diagrams not considered so far. The first diagram renormalizes both $\chi_{pp}$ and $\chi_{ph}$ alike, thus it cannot lead to a non-zero OF solution. The second ``rainbow" diagram is already contained in the self-consistent Eliashberg equation. The third diagram can contribute in principle, but it depends on $\omega + \omp$ even for frequency-independent interactions, and its contribution does not vanish for $\omega = \omp$. Thus, it cannot be treated in the Eliashberg framework.

\subsection{Cancellation of the thermal terms}

As discussed in the main text, at a finite temperature we can isolate two thermal contributions from the second-order self-energy diagram. However, we need to subtract the contribution shown in Fig.\ \ref{beyondEexpra}(b), where the frequency transfer on both lines is zero. But this contribution vanishes: To see this, we can expand the fermionic dispersion around the external momentum $\k$ as
\begin{align}
\xi(\k + \boldsymbol{p}) \simeq \xi(\k) + v_F p_{\parallel} + \boldsymbol{p}_{\perp}^2/(2m) ,
\end{align}
where $p_\parallel, \p_\perp$ are the components of $\p$ parallel and perpendicular to $\k$, respectively, and  $m$ is an effective mass. Expanding  $\xi(\k + \boldsymbol{p}+ \boldsymbol{l})$ in the same way, for instance the integral over $p_{\parallel}$ vanishes for zero frequency transfer: if the integral is evaluated by contour integration in infinite limits, both poles are in the same half-plane. Note that such an argument does not work if only one of the transferred frequencies is non-zero, while the other frequency is integrated over, since in this case the additional frequency integral must be evaluated before the momentum integrals, yielding non-zero.

As a result, the thermal contributions to $\Phi$ and $\Sigma$ in Eq.\ \eqref{eliashapp} read (with $\omega = \omp$ and after evaluating momentum integrals):
\begin{align}
\Phi_{\text{th}}(\omega) &= - \frac{\Phi(\omega)}{|\omega + \Sigma(\omega)|} {\rho \pi T V(0)}  \times \left[ 1 + 4\lambda f \right] \\  \notag  &= - \frac{\Delta(\omega)}{|\omega|}{\rho \pi T V(0)}\times \left [ 1 + 4\lambda f \right] \\  \notag
\Sigma _{\text{th}}(\omega) &= - \text{sign}(\omega) {\rho \pi T V(0)} \times  \left[ 1 + 4\lambda f \right] \\  \notag
\Delta_{\text{th}}(\omega) &= \Phi_\text{th}(\omega) - \frac{\Delta(\omega)}{\omega} \Sigma_\text{th}(\omega) = 0 \ .
\end{align}

\section{Behavior of $\lambda_c^e$}
\label{lambdaeApp}

\begin{figure}
\centering
\includegraphics[width=\columnwidth]{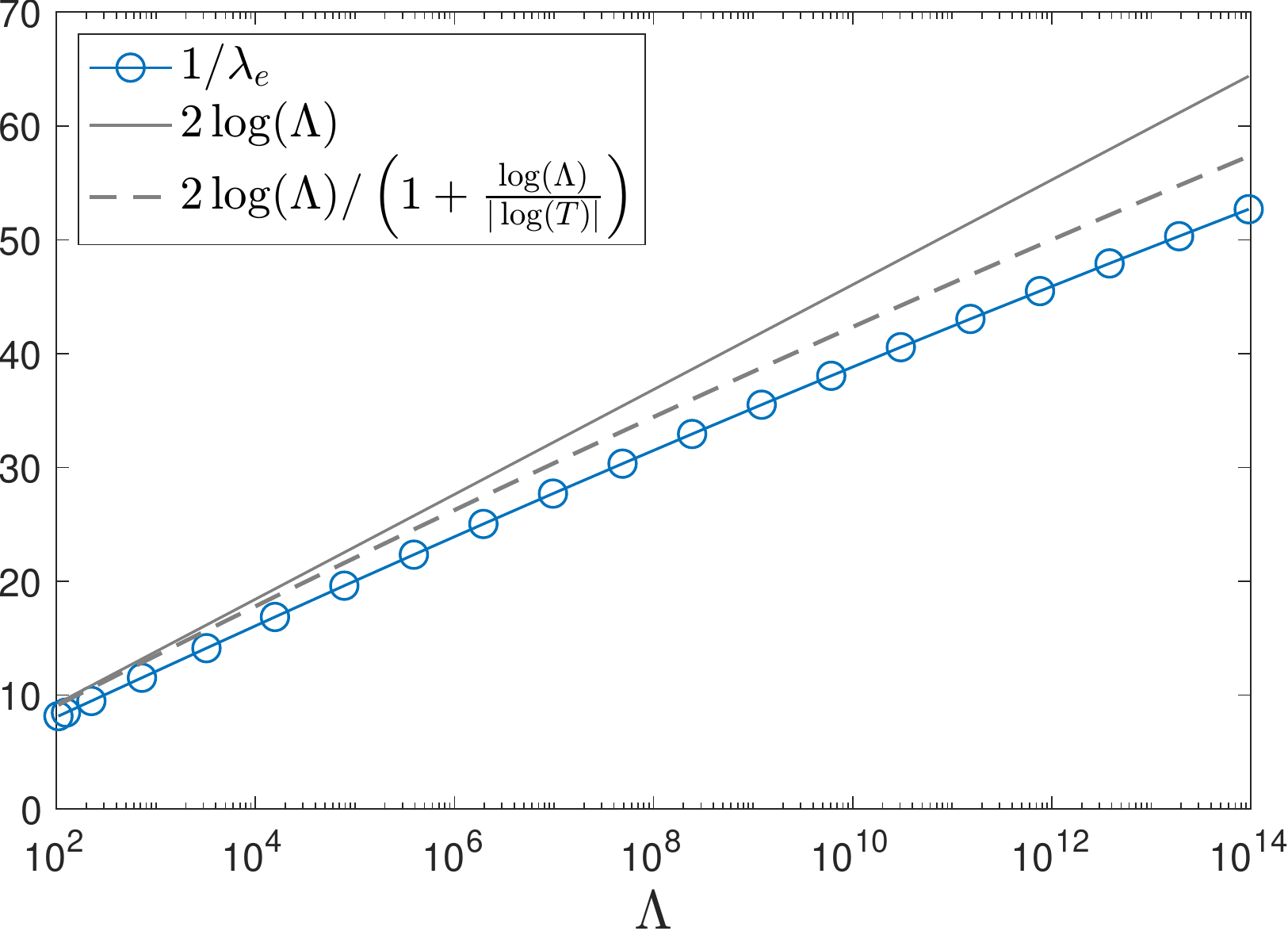}
\caption{Numerical estimation of $\lambda_c^e$ (blue line), without self-energy correction. Gray lines show the analytical estimates \eqref{lambdae} and \eqref{LambdaefiniteT}.}
\label{lambdaelargelambda}
\end{figure}

In Fig.\ \ref{lambdaelargelambda}, we numerically check the behavior of $\lambda_c^e \simeq 1/(2\log(\Lambda))$ at large $\Lambda$ by plotting $1/\lambda_c^e$. Apart from the nearly constant offset, which is expected since the formula only holds with ``logarithmic accuracy'', at very large $\Lambda$ the numerical result for $1/\lambda_c^e$ decreases compared to the asymptotic expression. This is expected since in the numerics a non-zero, though very small temperature $T$ is used, while $\lambda_c^e$ is defined as the critical coupling at zero temperature. Adapting the evaluation in Ref.~\cite{pimenov2021quantum} (see Eq.\ (15) within), one can provide an estimate for $\lambda_c^e$ at a finite temperature $T$ if one assumes that $T$ only serves as an IR cutoff, similar to $\Delta_e(0)$ in Ref.~\cite{pimenov2021quantum}. One finds
\begin{align}
\label{LambdaefiniteT}
\lambda_c^e = \frac{1}{2\log(\Lambda) }\times \left ( 1 + \frac{\log{\Lambda}}{|\log(T)|} \right).
\end{align}
As seen in Fig.~\ref{lambdaelargelambda}, this formula correctly reproduces the numerics up to the constant offset.

\section{Time-reversal transformation of the gap function}
\label{Ttrafoapp}

To derive the action of the time-reversal transformation $\T$ on gap functions on the Matsubara axis, we first derive the action on real-frequency gap functions. We work with retarded and advanced gap functions in the time domain at zero temperature, which are defined  as
\begin{align}
&\Delta^R(t) = - i \theta(t) \braket{0| \hat{M}(t) |0}  \\   \notag
&\Delta^A(t) = + i \theta(-t) \braket{0| \hat{M}(t) |0} ,
\end{align}
with $\ket{0}$ the interacting ground state, and
\begin{align}
\hat{M}(t) = \{ c_\alpha (t) , c_\beta(0) \} \left[i \sigma_y \cdot ({d_0} \mathbbm{1} + d_z \cdot  \sigma_z) \right]_{\alpha \beta} \ ,
\end{align}
where the spin term describes both singlet {$(d_0 = 1, d_z = 0)$} and mixed triplet  {$(d_0 = 0, d_z = 1)$}.
$\T$ acts on state vectors $\ket{\psi}$ and operators $\hat{O}$ as
\begin{align}
& \ket{ \T \psi} = T \ket {\psi}  \\  & \notag
\T \circ \hat{O} = T \hat{O} T^{-1} ,
\end{align}
where $T$ is an antiunitary matrix. Furthermore, expectation values fulfill
\begin{align}
\braket{\T \psi | \T \phi} = \braket{ \psi | T^\dagger T \phi} = \braket{ \psi | \phi}^{*} \ .
\end{align}
Using these properties, $\Delta^R$ transforms as
\begin{align}
& \left(\T \circ \Delta^R  \right )(t) = i \theta(-t) \braket{\T 0 |  \T \circ  \hat{M}(t)  |\T 0}  \\ & =  \notag
 i \theta(-t) \braket{ 0 | T^\dagger  T  \hat{M}(t) T^{-1} T   | 0} = i \theta(-t) \braket{0 |\hat{M}(t) | 0}^{*} \\
  & = - \left[\Delta^A(t) \right]^{*} \notag \ .
 \end{align}
Fourier-transforming, we obtain
\begin{align}
\label{realfreqtrafo}
& \left( \T \circ \Delta^R\right) (\Omega) =  \T \circ \int d t \exp(i\Omega t) \Delta^R(t)  =\\  \notag &  - \int dt \exp( - i\Omega (-t) )
\left [ \Delta^A(t) \right]^{*}  = - \left [  \Delta^A(-\Omega) \right]^{*} \  .
\end{align}
Note that $\T$ does not act on the measure $dt$, as can be checked considering the inverse Fourier transform. With Eq.\ \eqref{realfreqtrafo} at hand, we can infer the action of $\T$ on the Matsubara gap function $\Delta(\omega)$ from Cauchy's theorem. For concreteness, we focus on $\Delta_o$ and assume that $\Delta_o \in \mathbb{R}$ as in the main text. For $\omega > 0$, $\Delta_o$ is related to $\Delta_o^{R}$ and $\Delta_o^A$ as
\begin{align}
\label{Cauchy1}
&\Delta_o(\omega) = \frac{1}{2\pi i} \int d\Omega \frac{{\Delta_o^R(\omega)}}{\Omega - i \omega } \\ &
\label{Cauchy2}
\Delta_o(-\omega) = - \frac{1}{2\pi i} \int d\Omega \frac{\Delta_o^A(\Omega)}{\Omega + i \omega } \ .
\end{align}
By writing $\Delta^{R/A}_o(\Omega) = \Delta^{R/A}_1(\Omega) + i \Delta_2^{R/A}(\Omega)$, one can check that the condition $\Delta_o \in \mathbb{R}$ leads to
\begin{align}
\label{symDelta12}
&\Delta_1^{R/A}(\Omega) = \Delta_1^{R/A}(-\Omega)  \\  & \notag
\Delta_2^{R/A}(\Omega) = - \Delta_2^{R/A}(-\Omega) ,
\end{align}
In addition, from $\Delta_o(\omega) = - \Delta_o(-\omega)$ we obtain
\begin{align}
\Delta_o^{A}(\Omega) = - {\Delta_o^R(\Omega)^{*}} \ .
\end{align}
Now we can compute $( \T \circ \Delta_o)(\omega)$ for $\omega > 0$:
\begin{align}
&\left( \T \circ \Delta_o \right)(\omega) \overset{\eqref{Cauchy1},\eqref{realfreqtrafo}}{=}  - \frac{1}{2\pi i } \int d\Omega \frac{ - \left [  \Delta_o^A(-\Omega) \right]^{*} }{ \Omega + i\omega} \overset{\eqref{symDelta12}}{=}  \\   \notag & \frac{1}{2\pi i } \int d\Omega \frac{\Delta_o^A(\Omega)}{\Omega + i\omega}  \overset{\eqref{Cauchy2}}{=} - \Delta_o(-\omega) = \Delta_o(\omega) = \Delta_o(\omega)^{*} \ .
 \end{align}
Proceeding analogously for $\Delta_e \in \mathbbm{R}$, one also finds $(\T \circ \Delta_e) (\omega) = \Delta_e(\omega)^{*}$. Due to the antilinearity of $\T$, $(\T \circ \Delta)(\omega) = \Delta(\omega)^{*}$ then holds for arbitrary
$\Delta(\omega)$ of the form $\Delta(\omega) = \Delta_e(\omega) + \exp(i\phi) \Delta_o(\omega)$, as required in Sec.\ \ref{trsbsec}.
\\

{\section{Meissner effect}
\label{Meissnerapp}

The magnetic response of a superconductor is determined by the energy-cost of phase fluctuations. Given a spatially homogeneous mean-field solution $\Delta$ with fluctuating phase $\theta(\x)$, the momentum-space action for $\theta$ is
\begin{align}
\label{Stheta}
S_\theta \sim \int d\q \  n_s |\q|^2  \theta(\q) \theta(-\q) ,
\end{align}
where we neglected temporal fluctuations of $\theta$.  As shown in previous works \cite{PhysRevB.60.3485, PhysRevB.79.132502, doi:10.1143/JPSJ.80.054702,  RevModPhys.91.045005, PhysRevB.91.144514}, the superfluid density $n_s$, which enters Eq.\ \eqref{Stheta}, takes the same form for both EF and OF bulk superconductivity. We explicitly verified this result and confirmed it. At $T=0$, the superfluid density $n_s$, normalized to the normal-state density of electrons, is: 
\begin{align}
n_s = \frac{1}{2} \int d\omega \frac{|\Delta(\omega)|^2}{\left(|\Delta(\omega)|^2 + \omega^2 \right)^{3/2}} .
\end{align}
This is a manifestly positive expression. Therefore, the Meissner effect is diamagnetic in both EF and OF cases.

If we simply evaluate Eq.\ \eqref{Stheta} in the OF state at $T = 0$, we run into a problem: for $\Delta(\omega) \sim \omega$ as $\omega \rightarrow 0$ (see Sec.\ \ref{noselfsec}), the integral is logarithmically divergent at small frequencies, as already observed in Ref.\ \cite{PhysRevB.91.144514}. Note that this is the case only if the gap function is linear in $\omega$; for a general gap function which scales as $\Delta(\omega) \sim \omega^a$ with $a \neq 1$,  the integrand scales as
\begin{align}
\begin{cases}
\omega^{2a - 3}  \quad &a > 1 \\
\omega^{-a} \quad &a < 1 ,
\end{cases}
\end{align}
which leads to a convergent result.

For the given $\omega$-linear gap function, the logarithmic singularity of $n_s$ will be cut off by finite momenta $\q$. Therefore, the action for the phase field \eqref{Stheta} is modified to
\begin{align}
\label{Stheta}
S_\theta \sim \int d\q \  n_s(\q) |\q|^2  \theta(\q) \theta(-\q) , \quad n_s(\q) \sim \log(k_F/|\q|) \ .
\end{align}
The Meissner response can be obtained by coupling the system to an electromagnetic field $\boldsymbol{A}$. In the conventional case of Eq.\ \eqref{Stheta}, the constant $n_s$ acts a a mass term for $\boldsymbol{A}$, which leads to a penetration depth $\lambda \sim 1/\sqrt{n_s}$. Taken at face value, the logarithmic $n_s(\q)$ obtained in \eqref{Stheta} then implies a super-exponential decay of an external magnetic field in a superconductor, $B(x) \sim \exp(-x\log(x))$. However, we do not regard this as an observable effect, but rather as an artifact of the mean-field approximation: the logarithmic divergence of $n_s(\q)$ likely signals that in the full theory with fluctuations included, the gap function $\Delta(\omega)$ scales as $\omega^a$, with $a \neq 1$. As discussed above, in this case $n_s$  is finite, and the Meissner response is a conventional. }

\bibliography{repDelta}
\bibliographystyle{apsrev4-2}

\end{document}